\documentclass[letterpaper,journal]{IEEEtran}

\usepackage{etoolbox}
\providetoggle{original}
\settoggle{original}{false}
\usepackage[utf8]{inputenc}
\usepackage[T1]{fontenc}
\usepackage{citesort}
\usepackage{booktabs}
\usepackage{ragged2e}
\usepackage{amsmath,amsfonts}
\usepackage{algpseudocode}
\usepackage{array}
\usepackage{tabularx}
\usepackage[caption=false,font=scriptsize,labelfont=rm,textfont=rm]{subfig}
\usepackage{textcomp}
\usepackage{stfloats}
\usepackage{url}
\usepackage{verbatim}
\usepackage{graphicx}

\usepackage{amssymb}
\usepackage[ruled,vlined]{algorithm2e}

\begin{document}

\title{Analysis and Optimized CXL-Attached Memory Allocation for Long-Context LLM Fine-Tuning}

\author{
        Yong-Cheng Liaw, 
	Shuo-Han Chen\vspace{-0.1in}
}

\maketitle

\begin{abstract}\justifying
The substantial memory requirements of Large Language Models (LLMs), particularly for long-context fine-tuning, have renewed interest in CPU offloading to augment limited GPU memory. However, as context lengths grow, relying on CPU memory for intermediate states introduces a significant bottleneck that can exhaust the capacity of mainstream client platforms. To address this limitation, this work investigates the effectiveness of Compute Express Link (CXL) add-in card (AIC) memory as an extension to CPU memory, enabling larger model sizes and longer context lengths during fine-tuning. Extensive benchmarking reveals two critical challenges. First, current deep learning frameworks such as PyTorch lack fine-grained, per-tensor control over NUMA memory allocation, exposing only coarse, process-level policies. Second, due to this lack of control, when the memory footprint of fine-tuning is offloaded across local DRAM and CXL-attached memory, naively placing optimizer data in higher-latency CXL leads to substantial slowdowns in the optimizer step (e.g., $\sim$4× once data exceeds $\sim$20M elements). To overcome these challenges, this work introduces a PyTorch extension that enables tensor-level system memory control and a CXL-aware memory allocator that pins latency‑critical tensors in local DRAM while maximizing bandwidth by striping latency‑tolerant tensors across one or more CXL devices. Evaluated on a real hardware setup with 7B and 12B models, 4K--32K contexts, and a single GPU, CXL-aware allocation recovers throughput to 97--99\% of the local-DRAM baseline with a single AIC and matches the local-DRAM baseline with two AICs, delivering up to 21\% improvement over naive CXL while preserving DRAM‑like DMA bandwidth for GPU transfers. These results show that carefully managed CXL-attached memory is a practical path to scaling long-context fine-tuning beyond DRAM limits.
\end{abstract}


\vspace{-0.1in}
\section{Introduction} \label{S:Introduction}
The rapid growth of Large Language Models (LLMs) and their ever-increasing parameter counts have introduced significant challenges in memory capacity~\cite{minaee2025largelanguagemodelssurvey}. As these models frequently exceed available GPU memory, performance bottlenecks arise during both training and deployment. Memory requirements are further exacerbated by the push toward longer context lengths~\cite{liu2025comprehensivesurveylongcontext}, which arise from applications such as long chain-of-thought reasoning~\cite{deepseekai2025deepseekr1incentivizingreasoningcapability}, generative agents, in-context learning~\cite{geminiteam2024gemini15unlockingmultimodal}, retrieval-augmented generation~\cite{lee2024longcontextlanguagemodelssubsume}, and multimodal tasks~\cite{qwen2025qwen25vl}, all of which are experiencing rapid growth. To support these scenarios, fine-tuning LLMs on long-context datasets has become increasingly important~\cite{chen2024longloraefficientfinetuninglongcontext,an2024makellmfullyutilize,bai2024longwriterunleashing10000word,bai2024longalignrecipelongcontext}. However, long-context fine-tuning imposes substantial memory overhead, primarily from storing intermediate activation values that scale with context length~\cite{korthikanti2022reducingactivationrecomputationlarge,unsloth2025gradientcheckpointing}. Furthermore, limited memory resources restrict batch size during training, thereby constraining throughput.

To address these constraints, particularly in resource-limited environments, offloading strategies, such as CPU offloading and solid-state drive (SSD) offloading, have been proposed~\cite{ren2021zerooffloaddemocratizingbillionscalemodel,rajbhandari2021zeroinfinitybreakinggpumemory}. CPU offloading migrates model states, such as parameters, gradients, optimizer data, and occasionally checkpointed activations, from GPU memory to system memory (see Figure~\ref{fig:cpu_offloading_overview}), while SSD offloading further offloads these states onto SSDs, leveraging the larger and more cost-effective capacity of SSDs. However, SSD offloading suffers from inherent performance and endurance limitations of NAND flash memory, making CPU offloading the more practical and widely adopted approach~\cite{ren2021zerooffloaddemocratizingbillionscalemodel,huang2023elixirtrainlargelanguage,chen2025practicaloffloadingfinetuningllm,zeng2025autoheteautomaticefficientheterogeneous}. Although CPU offloading enables the fine-tuning of larger models and supports longer context lengths on GPU-memory-constrained systems, system memory capacity itself becomes the bottleneck instead, especially as model sizes continue to grow and the demand for longer contexts and larger batch sizes increases.

System memory capacity on mainstream client platforms (often ~192–256 GB today~\cite{amd20259950x3d}) is constrained by CPU/chipset limits, DIMM slots, and module density. To overcome these constraints, Compute Express Link (CXL) technology has emerged as a promising alternative, providing a viable solution to memory bottlenecks encountered during CPU-offloaded long-context LLM fine-tuning~\cite{cxlspec,chen2025nextgencomputingsystemscompute,wang2025exploringevaluatingrealworldcxl}. Leveraging PCIe and DRAM, CXL-attached memory expands host capacity beyond DIMM density/slot limits without the performance penalties of NAND-flash SSDs or the cost of high-capacity DIMMs~\cite{smartmcxlmemoryforlowertco}. In particular, CXL Type-3 add-in cards (AICs) provide memory expansion~\cite{microncxlmemory,smartmcxlmemoryforlowertco} that the host OS typically exposes as CPU-less NUMA nodes, allowing applications to access them similarly to remote DRAM, though with distinct performance characteristics. This capacity enables long-context LLM fine-tuning without being limited by the system memory size; however, since CPU offloading workloads are sensitive to system memory access latency, \textit{naively integrating CXL-attached memory into existing CPU offloading workflows does not inherently ensure optimal performance}.

Consequently, custom memory-management policies for CXL-attached memory tailored to offloading workloads are required~\cite{cxlanns,exploitingcxlbasedmemoryforddl}. Recent work has begun exploring CXL-attached memory for LLM workloads. For example, Wang et al.~\cite{wang2025exploringevaluatingrealworldcxl} evaluate end-to-end performance of CPU offloading with CXL, while Tang et al.~\cite{cxlmemoryexpansionforkvcachestorage} employ CXL-attached memory to store the KV cache during inference. However, these studies mainly characterize general performance without analyzing workload-specific behavior or proposing optimizations tailored to CPU-offloaded long-context LLM fine-tuning tasks. This leaves an open gap in understanding interactions such as frequent GPU–CPU transfers and latency-sensitive optimizer phases during long-context fine-tuning.

In practice, CPU offloading techniques such as ZeRO-Offload~\cite{ren2021zerooffloaddemocratizingbillionscalemodel} reduce GPU memory consumption by transferring model parameters, gradients, and optimizer states to system memory. While this effectively alleviates GPU pressure, it introduces frequent data transfers between GPU and CPU, and exposes performance sensitivity to memory access latency. When CXL-attached memory is used as an extension or replacement for local DRAM in offloading scenarios, its distinct performance characteristics, including higher latency and lower bandwidth compared to local DIMMs, need to be carefully considered~\cite{micronwhitepaperforcxlmemoryexpansion}. Generic operating system (OS)-level mechanisms, such as tiered memory systems~\cite{m52025masteringpagemigration,tpp2023} or interleaving policies~\cite{linuxv69numamemorypolicy}, provide transparent support but often yield suboptimal performance for specialized workloads. To clarify these implications, this work benchmarks CXL-attached memory under long-context CPU offloading workloads and identifies \textit{two primary performance challenges and one key performance characteristic.}

First, current deep learning frameworks such as PyTorch~\cite{paszke2019pytorch} enforce a single uniform allocation policy across DRAM and CXL-attached memory. Because memory is managed only at the process level, users are limited to coarse-grained tools like \texttt{numactl}~\cite{numactl_2025}, making fine-grained, per-tensor placement impossible and often leading to latency-sensitive tensors being allocated in CXL-attached memory. Second, CPU-based optimizer phases are particularly latency sensitive: read–modify–write loops over parameters, gradients, and optimizer states degrade significantly if data reside in CXL-attached memory rather than DRAM, as the memory accesses of the CPU-based optimizer step dominate the critical path and directly reduce throughput. On the other hand, this work identifies a key performance characteristic: in the GPU pipeline, asynchronous DMA overlaps data movement with kernel execution\footnote{A kernel is a low-level function that performs a specific tensor operation (e.g., convolution, matrix multiply). Kernel execution is running that function on hardware (CPU/GPU) to carry out the computation in parallel.}, and because both DRAM and CXL-attached memory traverse the same PCIe path, host-to-device transfer throughput is broadly comparable from either source.

To address these challenges and build on insights from prior analysis, this work introduces two primary optimizations. First, a fine-grained memory controller is designed and implemented as a PyTorch extension to enable per-tensor control over NUMA memory placement. Second, this paper introduces a CXL-aware memory allocator, implemented as a greedy policy that leverages the fine-grained memory controller to partition tensors by latency sensitivity. Latency-critical data are placed in local DRAM, while latency-tolerant data are directed to CXL-attached memory, with interleaving applied when beneficial to maximize bandwidth. In other words, through the introduced components, latency-bound optimizer states/updates in CPU-offloaded fine-tuning are pinned in DRAM, while bandwidth-bound GPU transfers are striped over CXL AICs via tensor-level placement. Together, these optimizations demonstrate that CXL-attached memory can effectively expand capacity for long-context LLM fine-tuning while achieving performance nearly identical to the local-DRAM baseline. Across 7B/12B models and 4K--32K contexts with a single GPU, CXL-aware allocation restores single-AIC throughput to 97--99\% of the local-DRAM baseline and matches the local-DRAM baseline with dual AICs, outperforming naive CXL by up to 21\%. The main contributions of this study are as follows.

\begin{enumerate}
    \item To the best of our knowledge, this paper presents the first empirical characterization of CXL-attached memory for long-context LLM fine-tuning, identifying and analyzing the key performance bottlenecks (e.g., optimizer latency sensitivity) introduced by naive CXL adoption.
    \item A PyTorch extension is implemented to enable precise, per-tensor memory placement on specified NUMA nodes, a crucial capability for heterogeneous memory systems.
    \item A CXL-aware memory allocator is introduced as a greedy policy that maps tensors by latency sensitivity and leverages interleaving when beneficial to maximize bandwidth and minimize latency-induced slowdowns.
    \item Real-world experimental results on CXL devices demonstrate that CXL-attached memory can expand capacity while delivering performance comparable to the local-DRAM baseline for long-context LLM fine-tuning workloads.
\end{enumerate}

The remainder of the paper is structured as follows: Section~\ref{sec:back} provides background and related work; Section~\ref{sec:analysis} analyzes CXL-attached memory performance; Section~\ref{sec:design} details our proposed optimizations; Section~\ref{sec:eval} presents experimental evaluations; Section~\ref{sec:related_works} reports related works and Section~\ref{sec:conclusion} concludes this study.

\vspace{-0.1in}
\section{Background}\label{sec:back}
This section provides background on CPU offloading techniques for long-context LLM fine-tuning (See Section~\ref{sec:back:cpu_offloading}), highlighting the associated system memory bottlenecks (See Section~\ref{sec:back:memory_scale}) and the role of Compute Express Link (CXL) technology in addressing these limitations (See Section~\ref{sec:back:cxl}).

\vspace{-0.1in}
\subsection{CPU Offloading for Long-Context Fine-Tuning}\label{sec:back:cpu_offloading}

ZeRO-Offload~\cite{ren2021zerooffloaddemocratizingbillionscalemodel} is a widely used technique to train LLMs on systems with limited GPU memory. It conserves GPU resources by transferring model parameters, gradients, and optimizer states from GPU memory to system memory, and only retrieves them back to GPU memory when required for computation. To further reduce memory usage, ZeRO-Offload can be combined with techniques such as Flash-Attention~\cite{dao2023flashattention2fasterattentionbetter}, Liger-Kernel~\cite{hsu2025ligerkernelefficienttriton}, and gradient checkpointing (activation checkpointing)~\cite{unsloth2025gradientcheckpointing}. Flash-Attention efficiently computes attention without fully materializing the attention matrix, ensuring that peak memory scales linearly rather than quadratically with context length. Liger-Kernel optimizes large intermediate tensor usage during cross-entropy calculation by employing a FusedLinearCrossEntropy mechanism. Notably, the intermediate tensor usage also scales with context length and vocabulary size. Activation checkpointing reduces peak memory by storing only a subset of activations during the forward pass and recomputing them during the backward pass. As context length grows, the volume of checkpointed activations increases, necessitating offload to system memory and on-demand retrieval~\cite{unsloth2025gradientcheckpointing}.

\begin{figure*}[!t]
    \centerline{\includegraphics[height=1.6in]{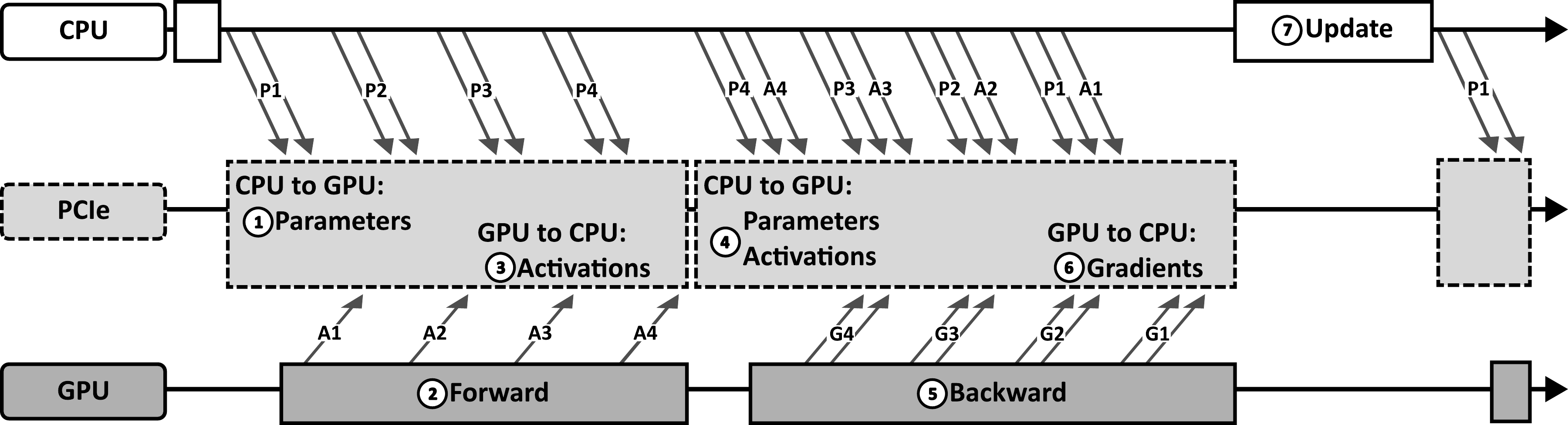}}
    \caption{Example of long-context CPU offloading with activation checkpointing with a transformer model composed of 4 transformer blocks. Arrows indicate data transfers over PCIe: $P_{i}$  represent model parameters (e.g., attention projection parameters, feed-forward network parameters) for a specific block. $A_{i}$ represents checkpointed input activations for the block. $G_{i}$ represents gradients corresponding to the parameters of the block. The numbered steps illustrate the data movement and computation flow.}
    \vspace{-0.15in}
    \label{fig:cpu_offloading_overview}
\end{figure*}

Figure~\ref{fig:cpu_offloading_overview} illustrates the aforementioned integrated approach, referred to as long-context CPU offloading with offloaded activation checkpointing, using a transformer model with four transformer blocks as an example. Each transformer block contains parameters for attention projections and feedforward layers. The workflow operates as follows: (1) First, necessary parameters are loaded from CPU to GPU memory on a tensor-by-tensor basis. (2) Next, the GPU performs forward computations using these parameters. (3) Checkpointed activations for each transformer block are offloaded to CPU memory. (4) Once the forward pass concludes, the backward pass requires parameters and previously checkpointed activations. (5) These data are then reloaded onto the GPU, which recomputes necessary activations to perform backpropagation. (6) Gradients computed on the GPU are subsequently offloaded to CPU memory, (7) enabling optimizer updates (e.g., using Adam) to execute entirely on the CPU after completing the backward pass. During optimizer steps, full precision parameters, optimizer states, and gradients reside primarily in CPU memory. Such an approach minimizes the volume of data transferred between the GPU and the CPU during each training iteration. Notably, the aforementioned workflow, which combines ZeRO-Offload, Flash-Attention, Liger-Kernel, and offloaded activation checkpointing, is considered the baseline.

To clarify the memory footprint of the workflow shown in Figure~\ref{fig:cpu_offloading_overview}, the GPU memory usage is first examined. During CPU offloading, the GPU is dedicated solely to computation and retains minimal data: model parameters are streamed block by block, and the corresponding activations and gradients are kept only until each block's computation is complete. These are then immediately offloaded to system memory or discarded, shifting the primary memory burden to the system memory. On the other hand, the system memory usage is detailed in Table~\ref{tab:memory_components}. The upper half lists components frequently transferred between CPU and GPU during forward and backward passes, while the lower half lists components stored on the CPU for optimizer updates. The memory usage for model parameters and gradients depends on their precision: \texttt{bf16} requires 2 bytes per parameter, while \texttt{fp32} requires 4 bytes per parameter. Notably, ZeRO-Offload uses \texttt{bf16} on GPUs to manage the huge memory footprint of activations and maximize throughput, while strategically using \texttt{fp32} on CPU for the sensitive optimizer calculations to ensure the model learns correctly and stably. For checkpointed activations, each GPU requires a unique set of activations; thus, the total system memory usage for these activations is scaled by $N_g$. Checkpoints are saved for each transformer block's input, totaling $L$ blocks, with each activation sized at $B \times C \times H$ elements, stored in \texttt{bf16} (2 bytes per element). For the Adam optimizer, the optimizer states (momentum and variance) require $8 \times P$ bytes in \texttt{fp32}, doubling the memory of gradients due to maintaining two states per parameter. Despite the aforementioned workflow having substantially reduced GPU memory usage, the workflow remains insufficient for long-context fine-tuning. This is because the memory required for activations grows enormously with sequence length. As a result, with parameters, gradients, optimizer states, and these massive activations all resident in system memory, memory pressure escalates rapidly, and system memory itself becomes the dominant bottleneck.

\begin{table}[htbp]
    \vspace{-0.1in}
    \centering
    \caption{System memory breakdown during CPU offloading. The upper half depicts CPU--GPU transfer size, and the lower half depicts system memory usage.
             $P$: total parameters; $N_g$: number of GPUs; 
             $B$: batch size per GPU; $C$: context length; 
             $L$: number of transformer blocks; $H$: hidden size.
    \vspace{-0.1in}}
    \label{tab:memory_components}
    \small
    \begin{tabularx}
    {\columnwidth}{@{} >{\RaggedRight}X c l @{}}
        \toprule
        Component                & Precision & Memory Usage (bytes)\\ \midrule
        Model parameters        & \texttt{bf16}      & $2 \times P$ \\
        Gradients               & \texttt{bf16}      & $2 \times P$ \\
        Checkpointed activations & \texttt{bf16}      & $2 \times (N_g \cdot B \cdot C \cdot L \cdot H)$ \\ \midrule
        Model parameters        & \texttt{fp32}      & $4 \times P$ \\
        Gradients               & \texttt{fp32}      & $4 \times P$ \\
        Optimizer states        & \texttt{fp32}      & $8 \times P$ \\
        \bottomrule
    \end{tabularx}
    \vspace{-0.15in}
\end{table}

\begin{figure*}[h]
    \centering
    \begin{minipage}{0.32\textwidth}
        \centering
        \includegraphics[height=1.5in]{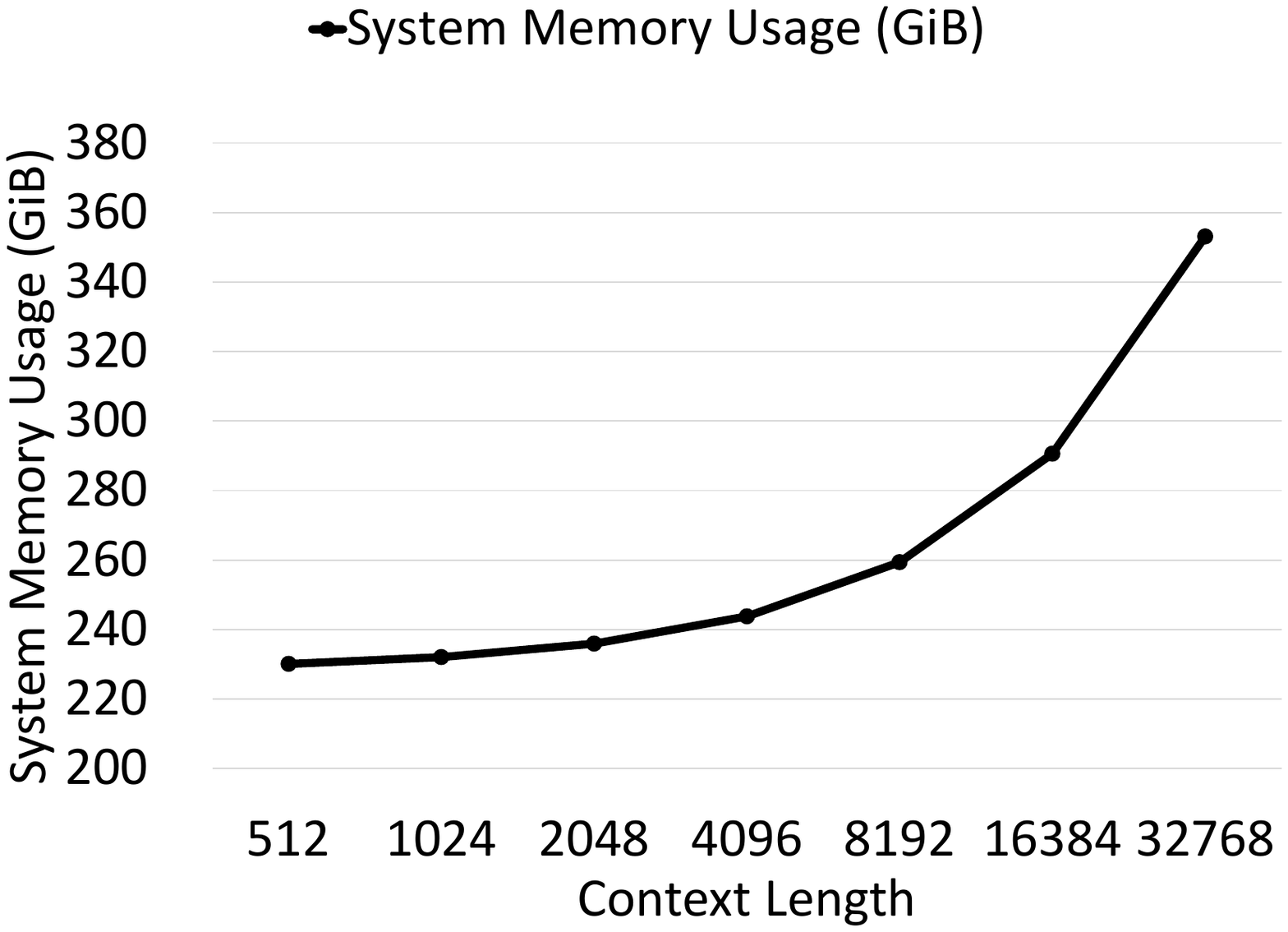}
        \caption{System memory requirement scaling for 12B across varying context lengths with a batch size of 5.}
        \label{fig:memory_usage_scaling_by_context}
    \end{minipage}%
    \hfill
    \begin{minipage}{0.32\textwidth}
        \centering
        \includegraphics[height=1.5in]{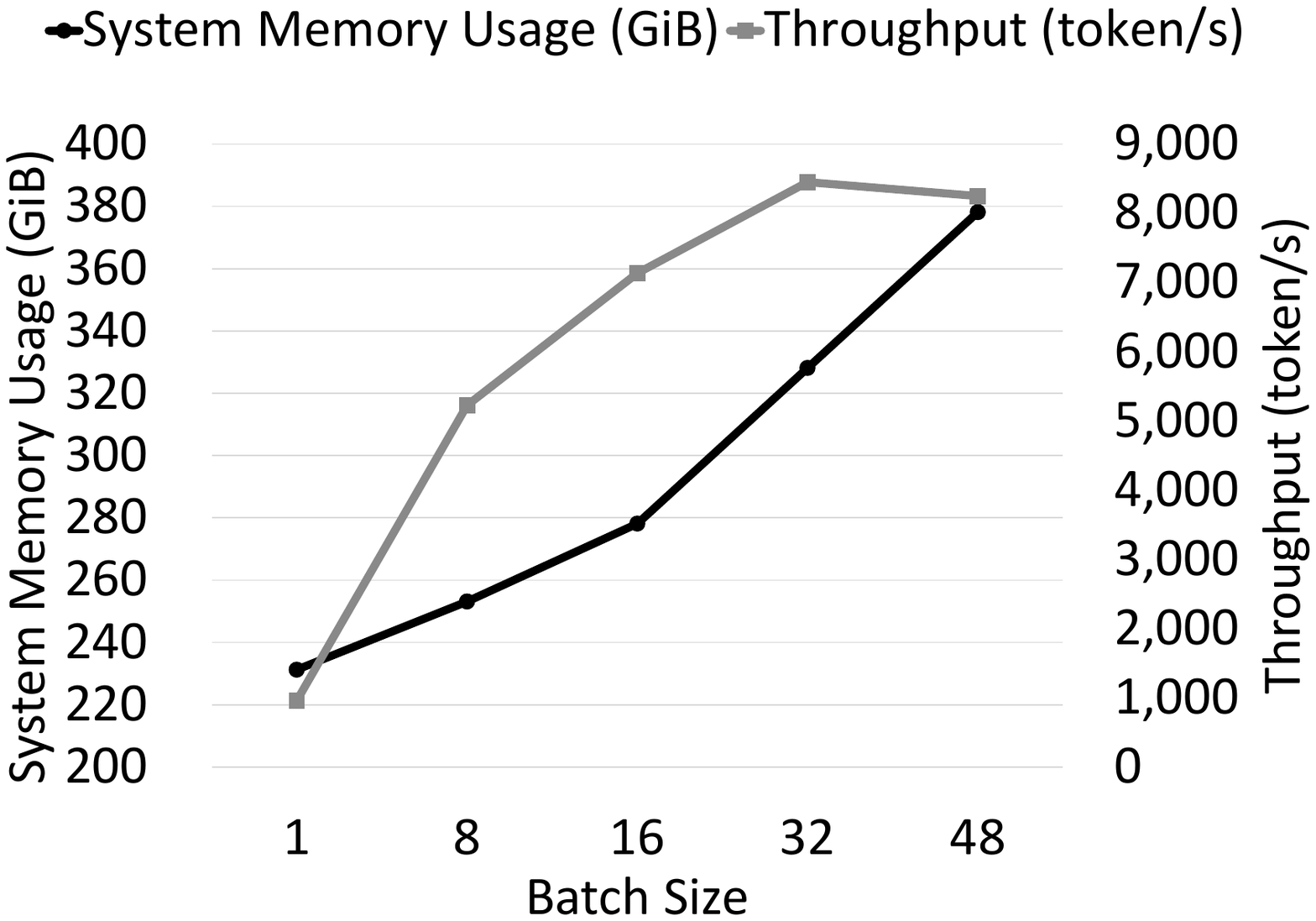}
        \caption{Throughput and system memory requirement scaling for 12B across batch sizes with a 4K context length.}
        \label{fig:memory_usage_scaling_by_batch}
    \end{minipage}
    \hfill
    \begin{minipage}{0.32\textwidth}
        \centering
    \includegraphics[height=1.5in]{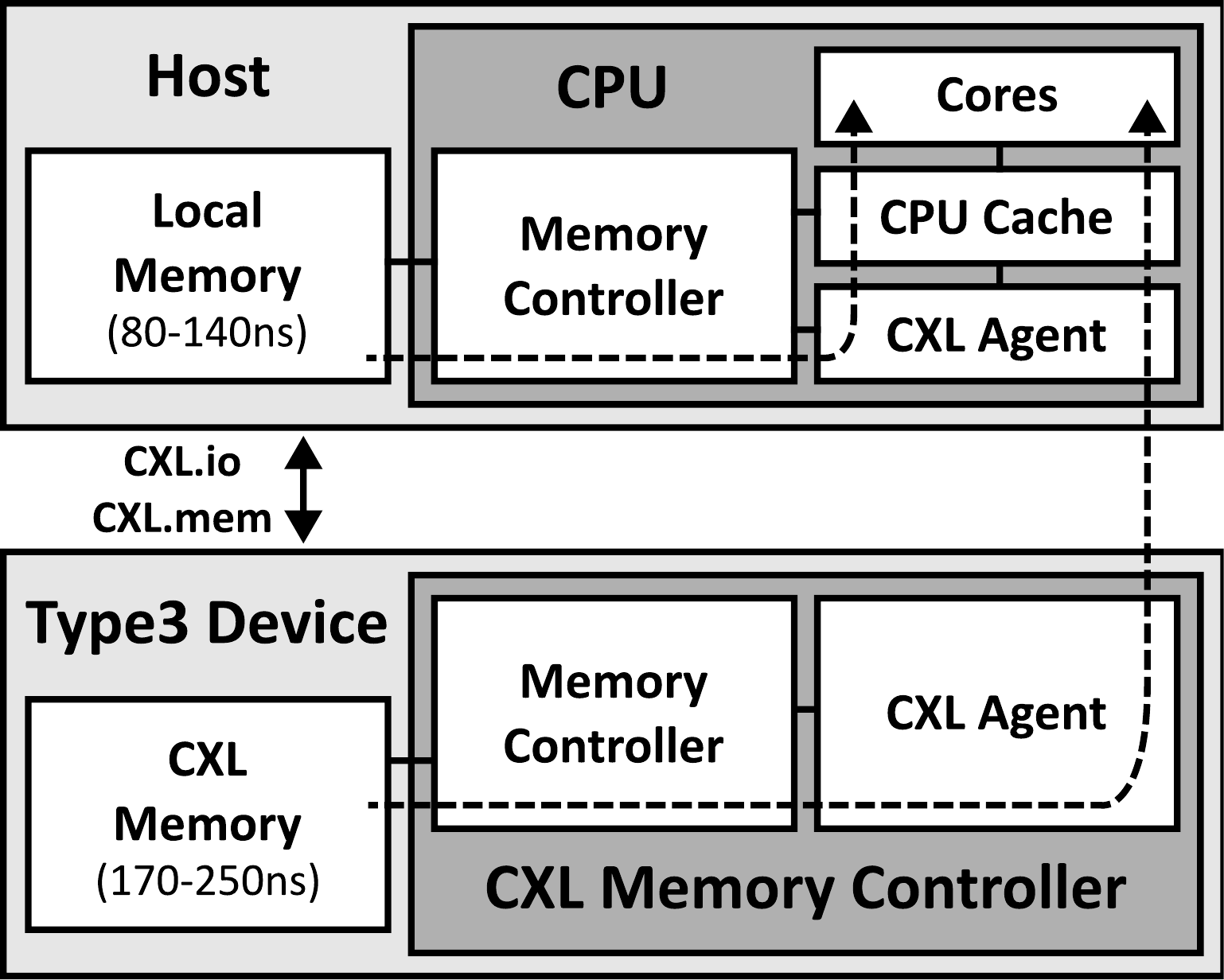}
    \caption{Comparison of memory access data paths and latencies between local memory and CXL-attached memory}
    \label{fig:cxl_memory_latency}
    \end{minipage}
    \vspace{-0.2in}
\end{figure*}

\subsection{Memory Bottleneck under Long-Context Offloading}\label{sec:back:memory_scale}

In the demanding scenario of training long-context LLMs with CPU offloading techniques, memory demand shifts predominantly from GPU memory to system memory. As a result, system memory capacity becomes a key factor, directly determining the feasible model size, maximum context length, and the batch sizes required to achieve optimal performance. 
To illustrate this behavior, a motivational experiment was conducted using the 12B model to measure memory requirements and throughput across different context lengths and batch sizes in a 2-GPU setting (hardware specifications are listed in Table~\ref{tab:system_spec}). In the first part of the experiment, the batch size is fixed at 5, and the context length is varied from 512 to 32K tokens. The choice of 32K is based on previous long-context fine-tuning studies~\cite{chen2024longloraefficientfinetuninglongcontext,an2024makellmfullyutilize,bai2024longwriterunleashing10000word,bai2024longalignrecipelongcontext}, which commonly use datasets with context lengths around 32K. For example, LongAlpaca~\cite{chen2024longloraefficientfinetuninglongcontext} ranges from 3K to 9K, FILM~\cite{an2024makellmfullyutilize} spans 4K to 32K, LongWriter~\cite{bai2024longwriterunleashing10000word} ranges from 2K to 32K, and LongAlign~\cite{bai2024longalignrecipelongcontext} ranges from 8K to 64K, with 90\% of samples below 32K. In the second part, the context length is fixed at 4K, while the batch size is varied from 1 to 48 to observe changes in throughput and memory usage. The results are presented in Figures~\ref{fig:memory_usage_scaling_by_context} and~\ref{fig:memory_usage_scaling_by_batch}.

Figure~\ref{fig:memory_usage_scaling_by_context} shows that CPU memory usage increases linearly with context length. This is because, in long-context CPU offloading, system memory needs to hold checkpointed activations, whose sizes scale proportionally with both the context length and the number of GPUs. Meanwhile, Figure~\ref{fig:memory_usage_scaling_by_batch} demonstrates that throughput improves with increasing batch size until saturation is reached. This suggests that once the model and context length are fixed, increasing the batch size can enhance GPU utilization. However, Figure~\ref{fig:memory_usage_scaling_by_batch} shows that CPU memory usage also increases linearly with batch size. This indicates that memory demand is driven not only by model scale and context length but also by batch size when aiming for optimal performance. These findings highlight that in long-context CPU offloading scenarios, system memory usage increases and is likely to become a critical bottleneck as context lengths continue to grow.

\begin{figure*}[h]
    \centering
    \begin{minipage}{0.32\textwidth}
        \centering
        \includegraphics[height=1.7in]{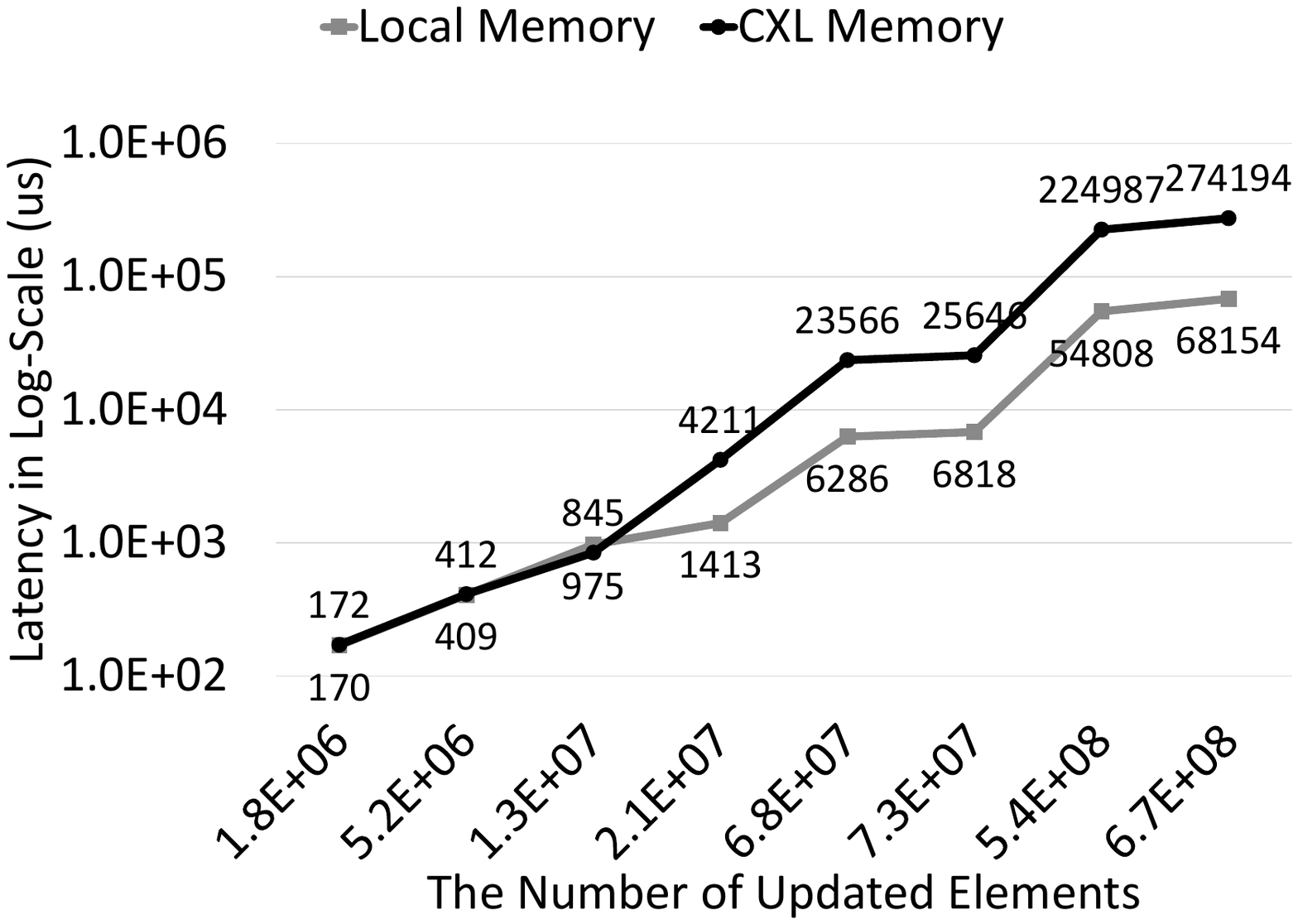}
        \vspace{-0.21in}
        \caption{Latency of the CPU-based Adam optimizer step with a growing number of parameters.}
        \label{fig:cpu_optimizer_bench}
    \end{minipage}%
    \hfill
    \begin{minipage}{0.29\textwidth}
        \centering
        \includegraphics[height=1.7in]{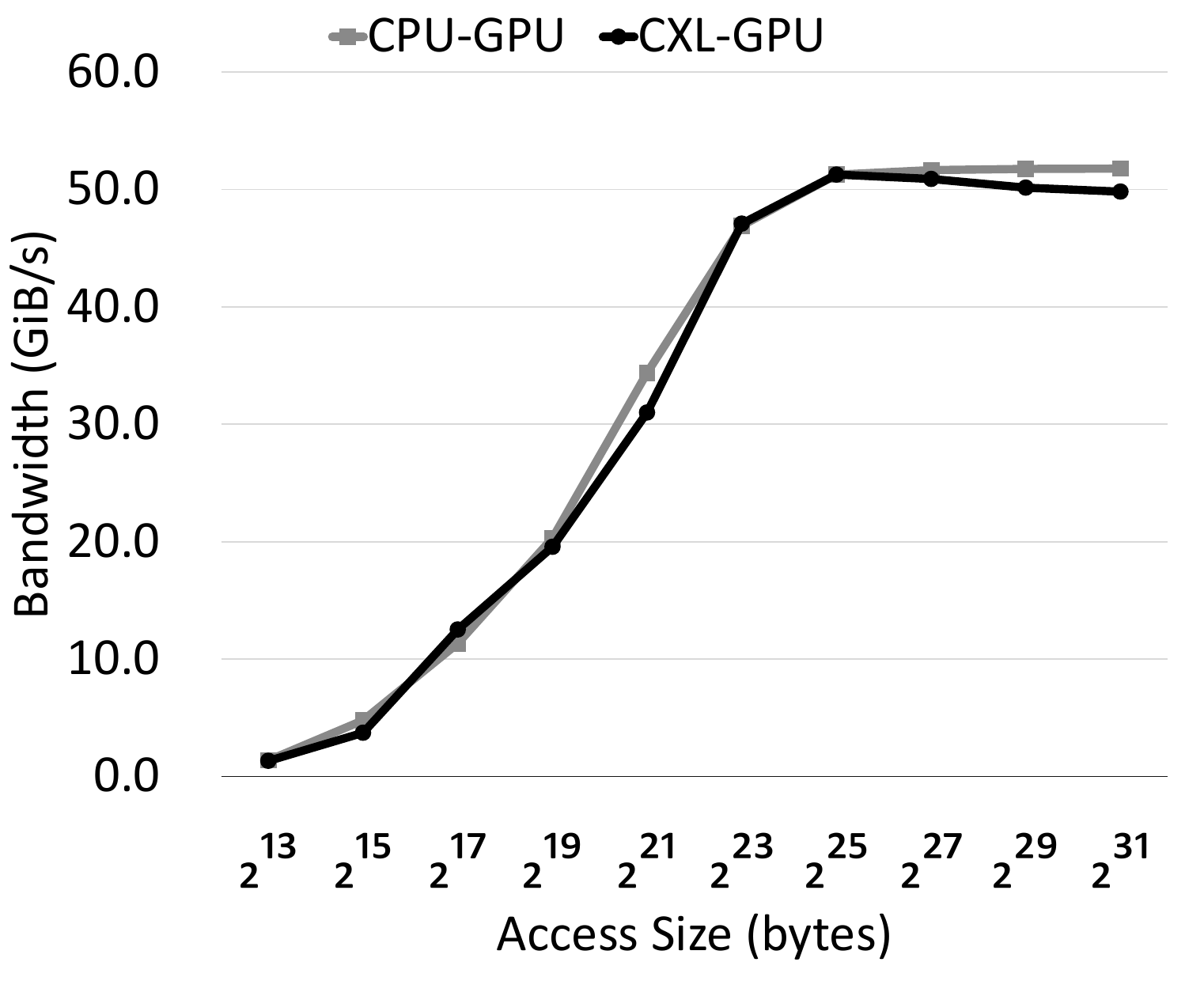}
        \vspace{-0.21in}
        \caption{Comparison of GPU transfer bandwidth from system DRAM and CXL-attached memory.}
        \label{fig:bandwidth_parity}   
    \end{minipage}
    \hfill
    \begin{minipage}{0.35\textwidth}
        \vspace{-0.2in}
        \centering
        \subfloat[Single CXL AIC]{\includegraphics[height=1.7in]{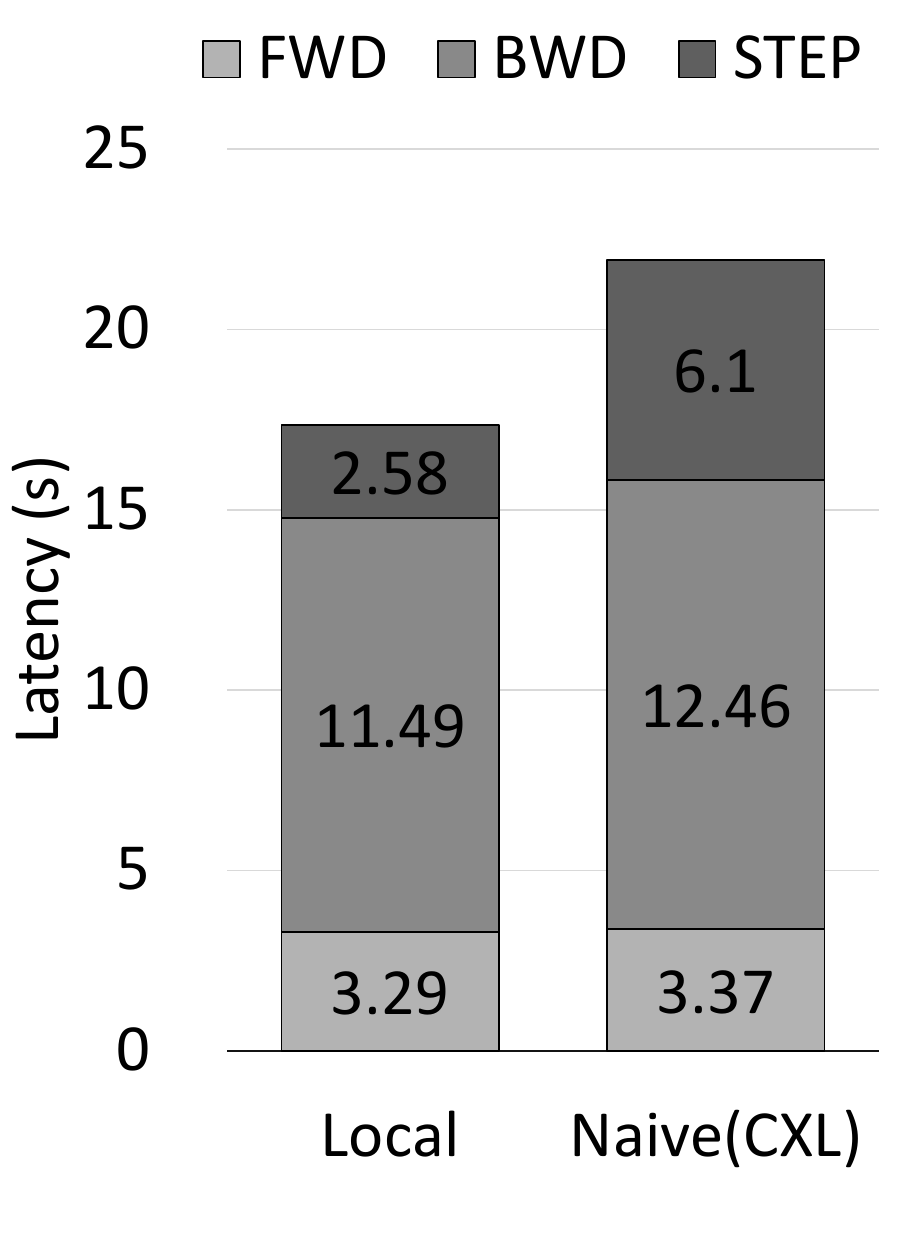}\label{fig:end_to_end_1gpu}}
        \subfloat[Dual CXL AICs]{\includegraphics[height=1.7in]{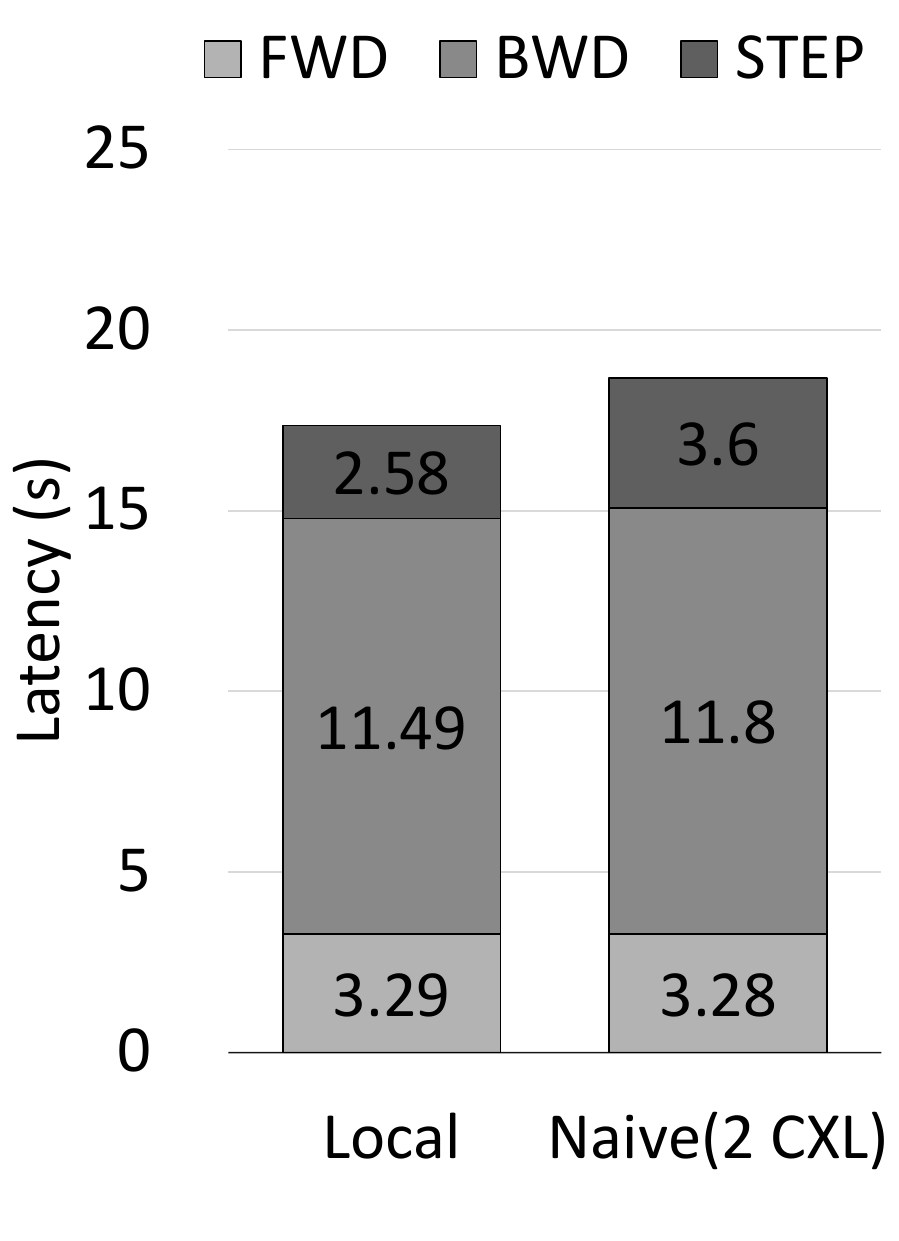}\label{fig:end_to_end_1gpu_2cxl}}
        \vspace{-0.05in}
        \caption{Latency breakdown of CPU offloading: local DRAM baseline vs. naive CXL configurations.}
        \label{fig:end_to_end_evaluation_for_motivation}
    \end{minipage}
    \vspace{-0.2in}
\end{figure*}

\vspace{-0.15in}
\subsection{CXL-Attached Memory}
\label{sec:back:cxl}
Compute Express Link (CXL) is built on top of PCIe and is designed to provide high-bandwidth, low-latency communication between the CPU (host) and various types of devices such as accelerators, memory expanders, and smart I/O devices. CXL differentiates devices into 3 types. Type 1 devices include accelerators with internal caches capable of directly caching host memory. Type 2 devices, such as GPUs, support mutual memory caching with the host, enabling unified memory access. Type 3 devices are intended for memory expansion and include components such as CXL-attached memory for expanding system memory capacity~\cite{cxlspec}. To enable the use cases of the above devices, three protocol sublayers, including CXL.io, CXL.cache, and CXL.mem, can be combined depending on the device type. While CXL.io provides traditional PCIe-like functionality for configuration, interrupts, and basic I/O operations, CXL.cache and CXL.mem enable devices to transparently cache host memory and allow the host to access memory attached to CXL devices, respectively.

For CXL-attached memory, Figure~\ref{fig:cxl_memory_latency} illustrates the differences in data paths and latency between local and Type 3 devices. Accessing local memory follows a direct path from the CPU cores through the CPU cache and memory controller, resulting in latencies between 80 and 140 nanoseconds~\cite{chen2025nextgencomputingsystemscompute}. In contrast, accessing CXL-attached memory involves traversing the PCIe interface using the CXL.io and CXL.mem protocols. This path requires coordination between the CPU and the CXL-attached memory controllers, leading to increased latency ranging from 170 to 250 nanoseconds~\cite{chen2025nextgencomputingsystemscompute}. While CXL-attached memory is utilized as a system memory extension, the Linux kernel integrates CXL-attached memory as CPU-less Non-Uniform Memory Access (NUMA) nodes~\cite{linuxv69numamemorypolicy}. This integration allows CXL-attached memory to be managed alongside traditional DRAM, while still enabling users to control allocation explicitly, such as through \texttt{numactl} or \texttt{libnuma}~\cite{numactl_2025}, to direct specific data to DRAM, CXL-attached memory, or an interleaved round-robin fashion among available NUMA nodes~\cite{numactl_2025}. Although CXL integration expands usable system memory, our analysis shows that CPU-offloaded long-context LLM fine-tuning remains suboptimal when backed by CXL-attached memory under current deep-learning frameworks (see Section~\ref{sec:analysis}).

\vspace{-0.1in}
\section{Observation and Analysis}\label{sec:analysis}
This section empirically characterizes the performance impact of storing CPU-offloaded data on CXL-attached memory. It begins by identifying a critical limitation in current deep learning frameworks that prevents fine-grained memory control. Subsequently, it analyzes the performance of distinct workload components, such as CPU-based computations and GPU data transfers, when using CXL-attached memory. The analysis reveals that naive CXL integration leads to significant end-to-end performance degradation. These findings collectively establish the motivation for the CXL-aware data placement strategies detailed later in this study.

\vspace{-0.15in}
\subsection{PyTorch Memory Allocation Limitation}\label{sec:analysis:pytorch_limitation}
PyTorch employs a layered execution stack that lowers high-level Python operations to efficient C++ back-end routines responsible for memory allocation and data movement. This architecture works well in homogeneous memory systems; however, it exposes a central limitation in heterogeneous settings with CXL-attached memory. That is, the current memory placement is controlled at the \textit{process level}, not at the granularity of individual tensors. In current deployments, NUMA policies configured externally (e.g., via \texttt{numactl}) are applied uniformly to the entire Python process. For example, launching a script with \texttt{numactl --interleave=0,1} causes \emph{all} tensor allocations in that process to be interleaved across NUMA nodes 0 and 1, regardless of their role or access pattern. Because PyTorch provides no native mechanism to annotate tensors with placement hints or to route specific allocations to distinct memory tiers, users cannot express policies such as “pin latency-critical optimizer states in local DRAM” while “spilling bandwidth-tolerant or cold tensors (e.g., checkpointed activations) to higher-capacity CXL-attached memory.”

This mismatch is particularly limiting for CPU-offloaded pipelines, where different tensor classes exhibit markedly different sensitivity to latency and bandwidth. A single, process-wide policy forces heterogeneous data into a uniform treatment, obscuring opportunities to reduce stall time on the critical path while leveraging capacity elsewhere. The result is suboptimal end-to-end performance and wasted hardware flexibility in mixed-memory environments. These observations motivate a fine-grained allocation extension that (i) exposes per-tensor placement control, (ii) integrates with existing PyTorch allocators without disrupting user code, and (iii) enables principled policies that align tensor characteristics with the most suitable memory tier.

\vspace{-0.15in}
\subsection{CPU Offloading Slowdowns on CXL-Attached Memory}\label{sec:analysis:cpu_compute}

As illustrated in Section~\ref{sec:back:cpu_offloading}, the
CPU-offloaded workflow stores full‑precision parameters, gradients, and optimizer states in system memory so that the CPU can perform the optimizer update locally. As each parameter update is independent, the optimizer phase exhibits ample parallelism. Practical implementations, such as ZeRO-Offload, exploit OpenMP threads and SIMD instructions (e.g., AVX2) to accelerate this compute-intensive step~\cite{ren2021zerooffloaddemocratizingbillionscalemodel}. Consequently, the optimizer phase is highly parallelized and sensitive to increased latency when accessing offloaded data structures. To quantify how memory placement affects optimizer latency within the long-context CPU-offloaded workflow, the CPU‑based Adam optimizer is benchmarked with offloaded data structures residing either in local DRAM or in CXL-attached memory. Figure~\ref{fig:cpu_optimizer_bench} summarizes the results. For each configuration, the data size is varied to emulate different LLM scales. Hardware details are listed in Table~\ref{tab:system_spec}, Config. A.

In Figure~\ref{fig:cpu_optimizer_bench}, an “element” consists of a 4-byte parameter, a 4-byte gradient, and 8 bytes of optimizer state. The SIMD kernel processes each element in three steps: (i) it loads the parameter, gradient, and state from memory into vector registers; (ii) executes the floating‑point update; and (iii) writes the updated values back. For modest data volumes, the latency penalty of CXL-attached memory is negligible; however, once the element count exceeds roughly 20 million, optimizer time with CXL-attached memory rises sharply, reaching nearly 4 times the DRAM baseline. The primary cause is the higher access latency of the CXL path (170-250 ns) compared to local DRAM (80-140 ns), as shown earlier in Figure~\ref{fig:cxl_memory_latency}. These results indicate that naively placing latency‑critical optimizer data in CXL-attached memory can severely degrade fine‑tuning performance. Effective CXL deployments for long-context CPU offloading need to keep latency‑sensitive data in low‑latency DRAM and relegate latency‑tolerant data to CXL-attached memory, respectively.

\vspace{-0.15in}
\subsection{GPU Data Transfers on CXL-Attached Memory}\label{sec:analysis:gpu_transfer}
The CPU offloading workflow involves not only a CPU-intensive optimizer step but also frequent, high-volume data transfers between system and GPU memory. For example, before each layer’s backward computation, model parameters and checkpointed activations are copied from system memory to the GPU. Afterward, the resulting gradients are copied back. To quantify the impact of physical memory location on data transfer behavior, this study evaluates GPU copy performance with source buffers placed in either local DRAM or CXL-attached memory. In each experiment, page-aligned host buffers are allocated and pinned to a specific NUMA node. These buffers are registered to enable direct DMA transfers over PCIe, thereby bypassing intermediate copies through CPU caches. Asynchronous memory transfers are then issued, and the resulting effective bandwidth is measured. Figure~\ref{fig:bandwidth_parity} shows the results.

\begin{figure}[b]
    \vspace{-0.15in}
    \centering
    \includegraphics[width=2.4in]{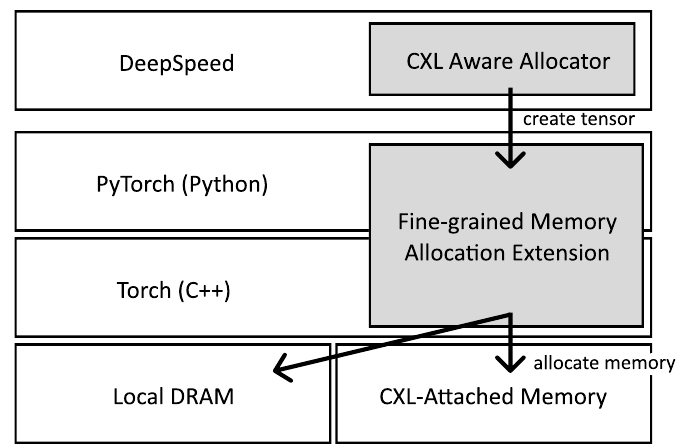}
    \vspace{-0.1in}
    \caption{System architecture of the proposed methods, in which a fine-grained allocation extension adds per-tensor placement control and a CXL-aware memory allocator directs tensors to local DRAM or CXL-attached memory within the existing software stack.}
    \label{fig:system_architecture}
\end{figure}

As shown in the figure, the observed transfer bandwidth from CXL-attached memory closely matches that of local DRAM. This is attributed to the use of direct DMA over PCIe, which allows CXL-attached memory to bypass the CPU and transfer directly to the GPU. Since local DRAM also relies on PCIe for such transfers, both memory types share a similar topology. Throughput increases with transfer size until it saturates the PCIe interface bandwidth. This convergence occurs because page-locked buffers expose equivalent DMA paths across both memory types, making the operation bound by interface limits. Furthermore, GPU data transfers tend to tolerate latency more effectively. This is because, for example, CPU offloading systems can utilize prefetching or asynchronous offloading to mask latency. These results indicate that for high-throughput, latency-tolerant operations such as GPU transfers, CXL-attached memory can deliver performance comparable to local DRAM. This contrasts with latency-sensitive phases, such as optimizer steps, which reinforce the importance of workload-aware data placement.

\vspace{-0.15in}
\subsection{End-to-End Fine-Tuning Slowdown}\label{sec:analysis:motivation}
Based on the characteristics outlined in previous sections, this section measures how naively incorporating CXL-attached memory affects end-to-end performance during LLM fine-tuning. The comparison evaluates a local-DRAM baseline against configurations that combine local DRAM with CXL-attached memory under a naive CXL policy. This naive CXL configuration is representative of what occurs due to the PyTorch memory allocation limitations described in Section~\ref{sec:analysis:pytorch_limitation}. Figure~\ref{fig:end_to_end_evaluation_for_motivation} presents the latency profile for fine-tuning a 12-billion-parameter model. In Figure~\ref{fig:end_to_end_evaluation_for_motivation}(a), a single CXL AIC is used. The optimizer step, labeled as the STEP phase, suffers the most significant slowdown. This is because its CPU-bound loads and stores are acutely sensitive to the higher access latency of CXL-attached memory, a direct consequence of the naive CXL policy placing critical optimizer data on the slower tier. Phases dominated by GPU transfers, specifically FWD and BWD, exhibit smaller slowdowns because prefetching and asynchronous DMA obscure some of the added latency.

Figure~\ref{fig:end_to_end_evaluation_for_motivation}(b) shows the result of adding a second CXL AIC. With more available bandwidth from the additional device, the overall performance improves, and the slowdown is mitigated compared to the single-AIC case. However, a performance gap still remains relative to the local-DRAM baseline. This demonstrates that while adding more hardware can alleviate bandwidth issues, it does not resolve the fundamental problem of latency sensitivity in the optimizer step. The naive CXL process-level memory policy remains a bottleneck. These findings underscore the need for a more intelligent CXL-aware allocation strategy. Such a strategy needs to distinguish between latency-sensitive and latency-tolerant data and leverage the aggregate bandwidth of multiple CXL AICs effectively.

\vspace{-0.1in}
\section{CXL-Aware Long-Context LLM Fine-Tuning}\label{sec:design}
To address the performance degradation arising from the naive adoption of CXL-attached memory, this work introduces a two-part methodology that overcomes the limitations of existing deep learning frameworks and intelligently manages data placement in heterogeneous memory systems. The first component is a fine-grained, per-tensor memory allocation extension for PyTorch, providing essential control over NUMA memory policies. The second is a CXL-aware memory allocator that leverages this extension to strategically place tensors based on their latency sensitivity. Figure~\ref{fig:system_architecture} illustrates the integration of these components into the LLM fine-tuning stack, where the extension operates between Python and C++ layers of PyTorch, and the CXL-aware allocator directs memory decisions from a higher level.

\begin{figure}[t]
    \centering
    \includegraphics[width=3.2in]{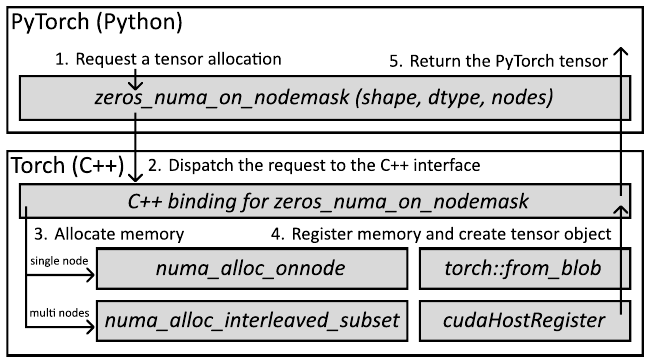}
    \caption{Workflow of the fine-grained memory allocation extension for PyTorch.}
    \label{fig:finegrained_extension}
    \vspace{-0.2in}
\end{figure}

\vspace{-0.1in}
\subsection{Fine-grained Memory Allocation Extension}
As discussed previously, on the path to exploiting CXL-attached memory under the CPU offloading scenario, a significant limitation of existing deep learning frameworks, such as PyTorch, lies in their coarse-grained memory management. By default, NUMA policies are enforced at the process level, causing all tensors allocated within a script to share the same placement strategy. This design prevents applications from selectively assigning tensors to different memory tiers according to their individual access patterns and latency sensitivities. To address this limitation, a fine-grained memory allocation extension for PyTorch is developed to introduce per-tensor control over NUMA placement, allowing developers to allocate tensors directly to local DRAM, a designated CXL device, or an interleaved set of nodes. The implementation is shown in Figure~\ref{fig:finegrained_extension}. It consists of a Python wrapper that interfaces with C++ backend functions and uses the \texttt{libnuma} library, specifically, \texttt{numa\_alloc\_onnode} for single-node placement and \texttt{numa\_alloc\_interleaved\_subset} for customized interleaving across nodes. To support efficient GPU data transfers, the extension leverages \texttt{cudaHostRegister} from the CUDA toolkit to pin host memory, thereby enabling zero-copy direct memory access (DMA), while \texttt{torch.from\_blob} is used to construct a PyTorch tensor that references the externally managed memory block without additional data copies. This extension establishes the foundational mechanism required for the CXL-aware memory allocator and enables future exploration of advanced memory management strategies in heterogeneous systems.

\vspace{-0.15in}
\subsection{CXL-Aware Memory Allocator}

To effectively exploit the aforementioned fine-grained, per-tensor control for the PyTorch memory allocation extension, this study further develops a CXL-aware memory allocator, which is an initialization-time placement algorithm that statically partitions CPU-offloaded data between local DRAM and CXL-attached memory. The allocator is designed to minimize performance degradation caused by CXL’s higher access latency by prioritizing data placement based on each component’s latency sensitivity. Its design builds on two key ideas: (1) hierarchical grouping of data by latency tolerance, (2) a latency-first greedy allocation strategy that enforces this hierarchy. In Section~\ref{demo}, the adaptive placement behavior is demonstrated under varying hardware configurations and memory pressure to demonstrate its effectiveness. Together, these mechanisms enable efficient and transparent utilization of heterogeneous memory resources during long-context LLM fine-tuning.

\subsubsection{Latency Sensitivity Classification}\label{sec:design:latency_sensitivity}

\begin{figure}[t]
    \centering
    \includegraphics[width=3.3in]{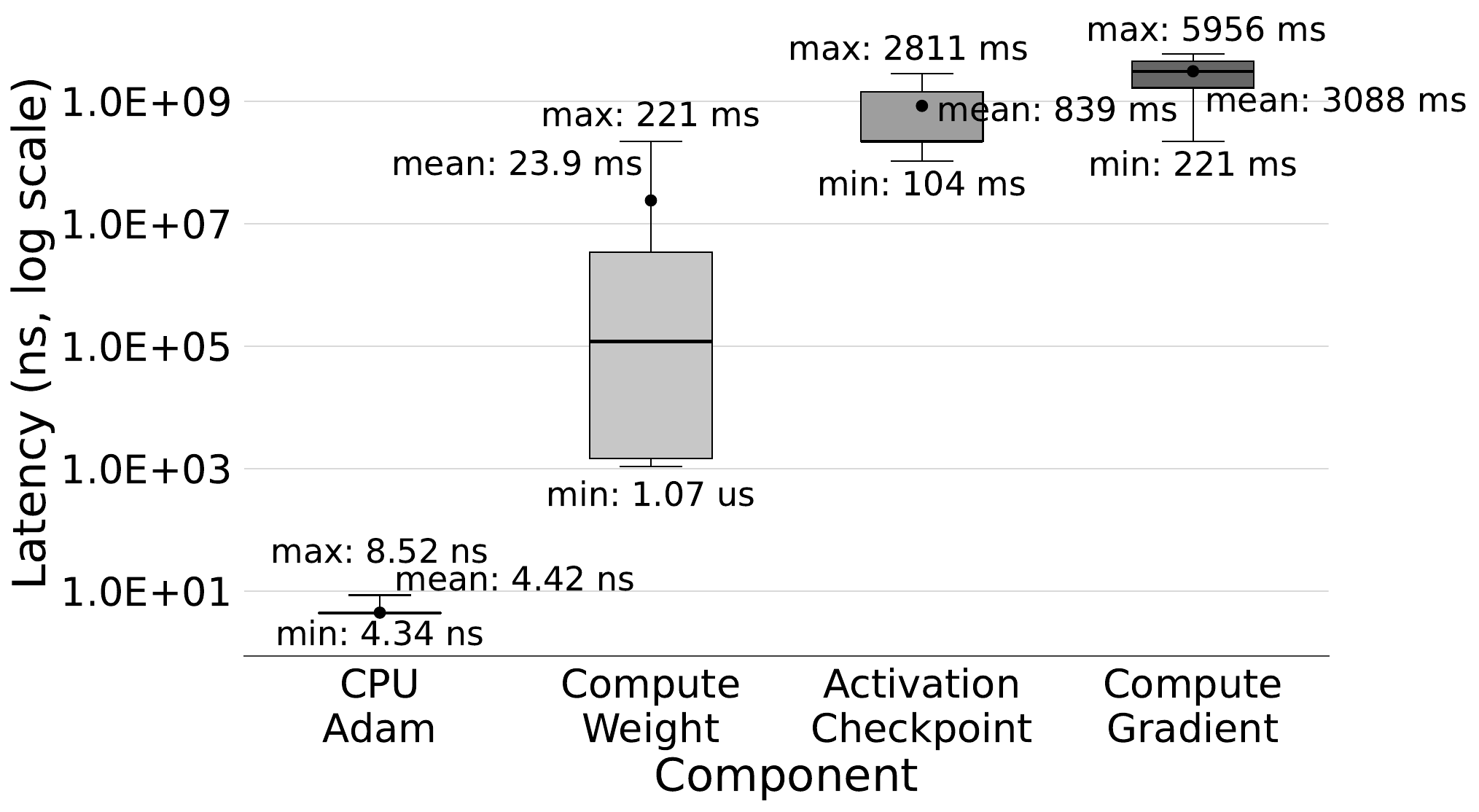}
    \vspace{-0.15in}
    \caption{Latency-sensitivity for CPU-resident components. A shorter latency indicates higher sensitivity.}
    \label{fig:latency_sensitivity_example}
    \vspace{-0.2in}
\end{figure}

As summarized in Table~\ref{tab:memory_components}, CPU-offloaded fine-tuning relies on system memory for several data components. To prioritize placement, the proposed allocator groups these components and assigns each a latency-sensitivity level. Figure~\ref{fig:latency_sensitivity_example} demonstrates how this sensitivity is derived from each component's budget based on their execution time or DRAM-based offloading latency. (for a 7B model, 32K context length, batch size 2, on the platform in Table~\ref{tab:ablation_system_spec}). This long-context setting is shown as a representative case, and the same four-level ordering holds across different context lengths and batch sizes. Tasks with minimal budgets, such as the nanosecond-scale CPU Adam, are severely bottlenecked by even minor memory delays. Therefore, latency sensitivity is inversely proportional to each component's budget. The distinct, orders-of-magnitude clustering seen in the figure justifies the multi-level classification scheme, where lower budgets dictate higher latency sensitivity and thus higher priority for local DRAM.

\textbf{Level~1} contains the data used only by the CPU Adam update, including \texttt{fp32} master parameters, gradients, and optimizer states. These tensors are accessed in tight load--compute--store loops on the CPU critical path via pipelined SIMD instructions.
Consequently, the effective tolerance to memory-access delay is bounded by the compute-only latency of a single SIMD chunk. In Figure~\ref{fig:latency_sensitivity_example}, this budget is below the 10~ns scale, making CPU Adam data the most latency-sensitive group. Notably, the remaining groups are more tolerant because their access is primarily from the GPU and can often be masked by prefetching or asynchronous offloading.

\textbf{Level~2} consists of \texttt{bf16} parameter computation required by each layer during forward and backward computation. The tolerance of layer prefetches is determined by the pure GPU computation time for one layer, which provides the slack to transfer parameters from CPU memory before the layer executes. However, the \texttt{bf16} parameters computation also requires a critical-path update after CPU Adam completes, in which the updated \texttt{fp32} master parameters are converted and copied to the \texttt{bf16} compute-parameter buffers. This copy path has only microsecond-scale tolerance, while layer prefetching has millisecond-scale tolerance, producing the wide range shown in Figure~\ref{fig:latency_sensitivity_example}. Consequently, compute parameters are less sensitive than CPU Adam data but still receive higher priority than purely transfer-overlapped data.

\textbf{Level~3} contains checkpointed activations. These tensors have two transfer patterns. After a layer finishes its forward computation, the generated activation can be asynchronously offloaded to CPU memory while later forward layers continue executing; the available slack for layer $i$ is therefore the remaining forward time from layer $i+1$ to the final layer. The second pattern occurs before the corresponding backward computation, where the activation must be prefetched back to GPU memory; this prefetch is constrained by the pure GPU backward time of one layer. Since activation prefetching can fall on the GPU critical path, checkpointed activations are assigned a tighter level than gradients, because they have a higher chance of stalling GPU execution if loading is not finished in time.

\textbf{Level~4} contains \texttt{bf16} compute gradients. After the backward computation of layer $i$ produces a gradient tensor, the tensor can be asynchronously offloaded while the backward computation proceeds through earlier layers. The available slack is the remaining backward time from layer $i-1$ to the first layer. Unlike compute parameters, these gradients require no additional format conversion, synchronization, or prefetch before the optimizer step; they are consumed only after the CPU Adam phase starts. This makes them the most latency-tolerant group.

Notably, the max/min latency difference in Levels~3 and~4 results from the layer-dependent asynchronous offloading budgets. Because a tensor's slack time spans from its generation until its next use, these budgets vary significantly. Early-checkpointed activations can hide offload latency behind all subsequent forward layers, while late-layer gradients, generated earliest during backward propagation, similarly have extended slack times.

\begin{algorithm}
\caption{Latency-First Allocation}
\label{alg:latency_first}
\KwIn{%
  $S_{\text{local}}$: Local DRAM size;\\
  $S_{\text{cxl}}$: Aggregated CXL size;\\
  $num_{\text{cxl}}$: Number of CXL devices;\\
  \textit{group\_items\_sizes}: maps each latency level to its list of item sizes $\{level:[size_1,\dots]\}$}
\KwOut{\textit{allocations}: item $\rightarrow$ (policy, interleave ratio)}

$S_{\text{remain}} \leftarrow S_{\text{local}}$\;

\textit{allocations} $\leftarrow \{\}$\;

\For{$level \leftarrow 1$ \KwTo $4$}{
  \ForEach{$size_i$ \textbf{in} $group\_items\_sizes[level]$}{
    $key \leftarrow$ "\texttt{level}\_$level$\_\texttt{item}\_$i$";
    
    \uIf{$S_{\text{remain}} \ge size_i$}{
      \textit{allocations}[$key$] $\leftarrow$ (\texttt{pure\_local}, $1{:}0$)\
      
      $S_{\text{remain}} \leftarrow S_{\text{remain}} - size_i$\;
    }
    \uElseIf{$S_{\text{remain}} > 0$}{
      \textit{allocations}[$key$] $\leftarrow$ (\texttt{local\_cxl},
            $(1{:}\underbrace{1{:}\dots{:}1}_{num_{\text{cxl}}})$)\;

      $S_{\text{local\_used}} \leftarrow
            \min\left(\left\lfloor \frac{size_i}{1+num_{\text{cxl}}} \right\rfloor,
            S_{\text{remain}}\right)$\;

      $S_{\text{remain}} \leftarrow S_{\text{remain}} - S_{\text{local\_used}}$\;
    }
    \Else{
      \textit{allocations}[$key$] $\leftarrow$ (\texttt{pure\_cxl},
            $\underbrace{1{:}\dots{:}1}_{num_{\text{cxl}}}$)\;
    }
  }
}
\Return \textit{allocations}\;
\end{algorithm}
\setlength{\textfloatsep}{5pt}%

\begin{figure*}[t]
    \centering
    \begin{minipage}{0.25\textwidth}
        \centering
        \includegraphics[height=1.3in]{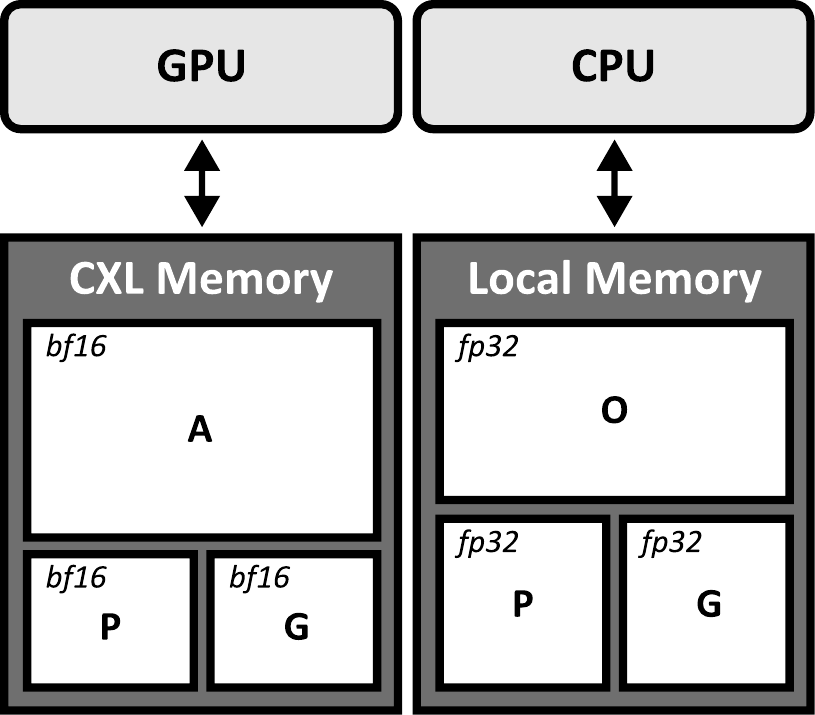}
        \caption{Allocation example for single CXL device and sufficient DRAM.}
        \label{fig:method_example_1}
    \end{minipage}\hfill
    \begin{minipage}{0.35\textwidth}
        \centering
        \includegraphics[height=1.3in]{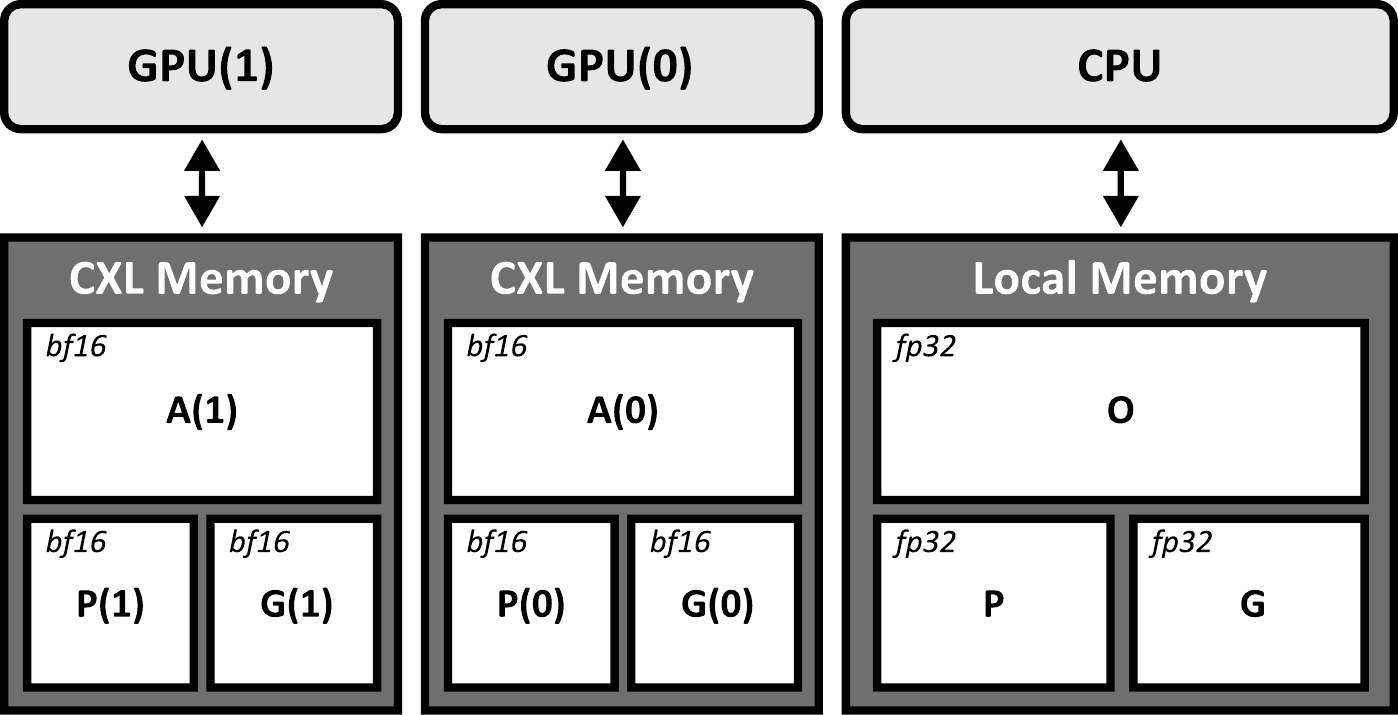}
        \caption{Allocation example for multiple CXL devices and sufficient DRAM.}
        \label{fig:method_example_2}
    \end{minipage}\hfill
    \begin{minipage}{0.35\textwidth}
        \centering
        \includegraphics[height=1.3in]{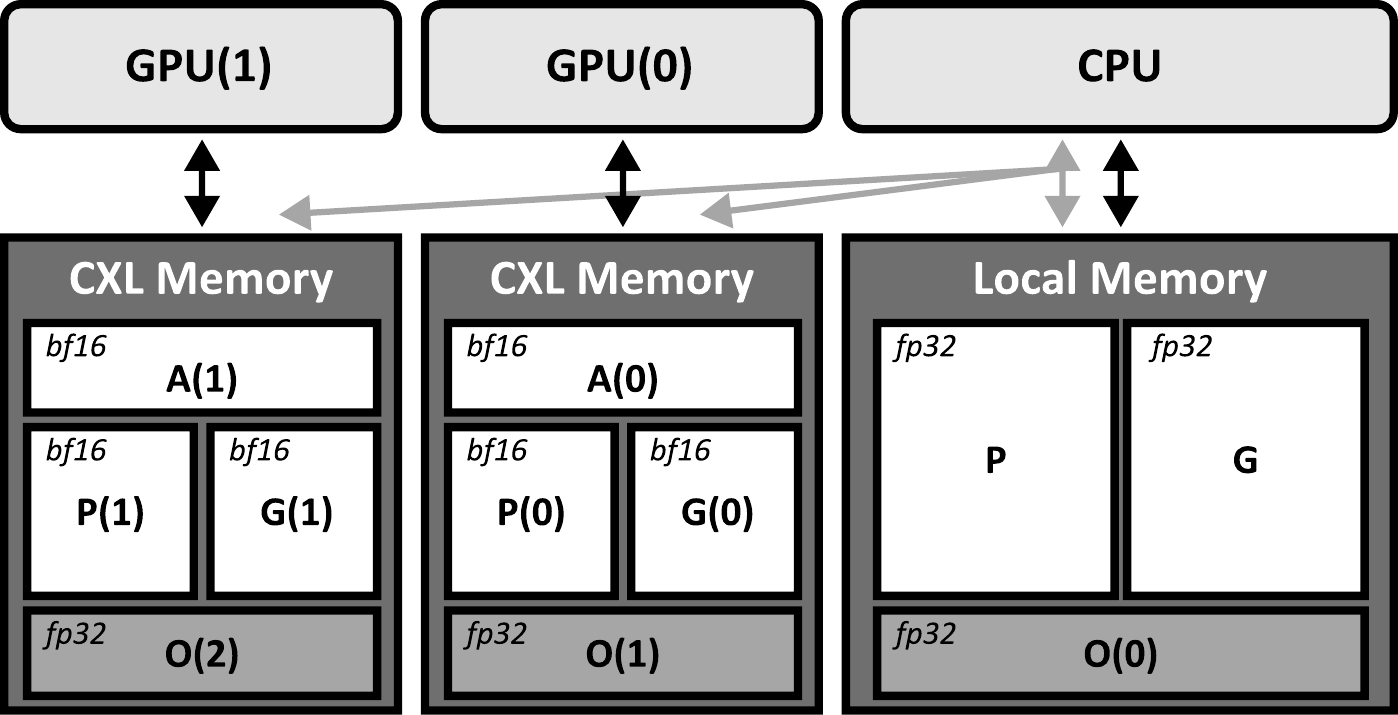}
        \caption{Allocation example for multiple CXL devices and limited DRAM.}
        \label{fig:method_example_3}
    \end{minipage}
    \vspace{-0.15in}
\end{figure*}

\subsubsection{Latency-First Greedy Algorithm}
Building upon the four-level latency classification, the latency-first greedy algorithm determines how CPU-resident data are distributed across local DRAM and multiple CXL-attached memory devices. The guiding principle of the algorithm is to prioritize low-latency data placement in local DRAM while progressively offloading or interleaving more latency-tolerant data into CXL-attached memory as DRAM capacity becomes limited. This design follows a greedy strategy, which makes locally optimal decisions at each step, so that latency-critical data always receive preferential treatment without incurring the complexity of global optimization or iterative tuning. Algorithm \ref{alg:latency_first} presents the detailed allocation process. The algorithm takes as input the available local DRAM capacity ($S_{local}$), the aggregated capacity of all CXL devices ($S_{cxl}$), and the number of CXL devices ($num_{cxl}$). The CPU-resident data are grouped by latency sensitivity into the mapping structure \textit{group\_items\_sizes}, which associates each of the four tolerance levels with a list of tensor or parameter sizes. The output is an allocation table that maps each data item to a specific placement policy and its corresponding interleave ratio across memory tiers. The allocator begins with the full DRAM capacity ($S_{remain}$=$S_{local}$) and iterates through the latency levels from the most sensitive classification (Level 1) to the least sensitive classification (Level 4). Within each level, every item is evaluated based on its size and the remaining local DRAM space. 

If an item completely fits within the available DRAM ($S_{remain} \geq Size_{i}$), it is placed entirely in local DRAM, marked as \texttt{pure\_local}, ensuring minimal latency for the most performance-critical components such as optimizer states and master gradients. If the item cannot fit entirely while some DRAM space remains ($S_{remain}>0$), the allocator assigns the \texttt{local\_cxl} policy to interleave the item across local DRAM and all CXL devices using an even ratio. In this case, selecting \texttt{local\_cxl} does not directly exhaust the remaining DRAM budget; the allocator deducts the local allocation, $\min(\lfloor Size_i/(1+num_{cxl}) \rfloor, S_{remain})$, and places the rest of the item in CXL-attached memory. This placement uses local DRAM for part of the allocation while absorbing the overflow in CXL-attached memory and exploiting the aggregate bandwidth of the available CXL devices. Once DRAM is exhausted ($S_{remain}=0$), subsequent items are assigned the \texttt{pure\_cxl} policy and distributed evenly across all CXL devices to maximize capacity and aggregate bandwidth.

Notably, the latency-first greedy algorithm is static and deterministic, making it well-suited for CPU-offloaded fine-tuning where tensor access patterns remain largely unchanged across training iterations. In addition, CPU--GPU transfers rely on page-locked CUDA pinned memory to avoid extra host-side copies and enable efficient DMA transfers. Therefore, a static allocation strategy is preferable to dynamic migration or online tuning in this setting. Pinned memory is allocated and registered at the beginning of training and remains fixed throughout execution. Changing data placement at runtime would require reallocation and re-registration of pinned buffers, introducing substantial overhead. Alternatively, using unpinned memory could avoid re-registration costs, but it would introduce additional host-side copies during GPU transfers and therefore reduce performance.

Following the static nature of the latency-first greedy algorithm, placement is computed once from the model shape, context length, batch size, and available DRAM/CXL capacity, avoiding any need for online profiling, runtime migration, and iterative tuning. This enables the allocator to be implemented outside the existing system; it only redirects the allocation location during initialization, leaving the training loop, optimizer update, and offloading schedule unchanged. 
Despite this simplicity, the evaluation in Section~\ref{sec:eval} shows that the latency-first greedy algorithm consistently outperforms naive CXL across context lengths, batch sizes, optimizers, and model architectures. The method yields the highest performance gains when local DRAM cannot hold the full offloaded footprint but still holds most of the latency-sensitive optimizer data, provided that the aggregate CXL bandwidth remains sufficient for CPU--GPU transfers.

\subsubsection{Allocation Examples}\label{demo}
The behavior of the CXL-aware allocator is illustrated through three representative scenarios that reveal how the algorithm adapts to different hardware configurations and memory pressures. 

\begin{itemize}
    \item \textbf{Scenario 1 (Single CXL Device, Ample DRAM).} 
    As shown in Figure~\ref{fig:method_example_1}, a system equipped with one CXL device and sufficient DRAM places all latency-critical Level~1 data, which includes optimizer states (\texttt{fp32} O), master parameters (\texttt{fp32} P), and master gradients (\texttt{fp32} G), entirely in DRAM. The more latency-tolerant data, such as checkpointed activations (\texttt{bf16} A), parameters (\texttt{bf16} P), and gradients (\texttt{bf16} G), are stored in the CXL-attached memory.

    \item \textbf{Scenario 2 (Multiple CXL Devices, Sufficient DRAM).} 
    In Figure~\ref{fig:method_example_2}, multiple CXL devices are introduced while DRAM capacity remains sufficient for Level~1 data. Latency-sensitive data stay in DRAM, whereas latency-tolerant data (\texttt{bf16} A, \texttt{bf16} P, \texttt{bf16} G) are interleaved across the CXL devices to exploit their aggregate bandwidth, enhancing throughput for GPU data transfers.

    \item \textbf{Scenario 3 (Multiple CXL Devices, Limited DRAM).} 
    As illustrated in Figure~\ref{fig:method_example_3}, when DRAM is insufficient to accommodate all Level~1 data, the greedy allocator first fills the available DRAM with the highest-priority tensors. The remaining Level~1 data, such as part of the optimizer states (\texttt{fp32} O), are striped across the remaining DRAM and all CXL devices. This fallback mechanism ensures balanced utilization of memory resources while minimizing the performance penalties associated with higher-latency CXL access.
\end{itemize}

\vspace{-0.1in}
\section{Evaluations}\label{sec:eval}

\subsection{Experimental Setup}\label{sec:eval:setup}
\subsubsection{Hardware and Software Specification} \label{sec:eval:spec}
Experiments run on a server whose hardware and software stack are summarized in Table~\ref{tab:system_spec}. The platform combines a high‑performance CPU (e.g., Intel Xeon 6780E), ample local DRAM (e.g., 512 GB DDR5), and two cutting‑edge GPUs (e.g., NVIDIA H100) optimized for LLM workloads. CXL-attached memory expansion is evaluated in two configurations: a single-AIC setup and a dual-AIC setup designed to reveal scalability limits and bandwidth‑contention effects. Both add‑in cards were SMART Modular CMM-CXL-2.0 devices~\cite{smartmcxlmemoryforlowertco}. Notably, even though the current configuration is limited to the availability of CXL AICs, we argue that the observations made in this study are representative and transferable, as they are governed by the fundamental bandwidth and latency characteristics of the CXL protocol itself.

\subsubsection{Workload Setup}\label{sec:eval:workloads}
Data placement across local DRAM and CXL-attached memory is managed by \texttt{libnuma}, which is exposed to PyTorch through a lightweight custom extension that intercepts memory allocation calls to enforce NUMA policies. DeepSpeed~\cite{deepspeed}, which is the implementation of ZeRO-Offload, handles CPU offloading, while Flash-Attention, Liger-Kernel, and activation checkpointing enable efficient long‑context processing. Checkpointed activations, once generated, are offloaded to host DRAM. The study focuses on fine‑tuning large language models under a Causal Language Modeling objective. Two representative models, including \textbf{Qwen2.5-7B}~\cite{qwen2025qwen25technicalreport} and \textbf{Mistral NeMo 12B}~\cite{mistralnemo}, serve as workloads for exploring performance and scalability across varying context lengths, batch sizes, GPU counts, and AIC configurations. All experiments employ \texttt{bf16} mixed‑precision training. The Adam optimizer maintains \texttt{fp32} master parameters and optimizer states on the CPU, delivering the numerical stability required for LLM fine‑tuning. Unless otherwise specified, each reported result is computed as the average over five independent runs. Details on specific context lengths, batch sizes, and AIC setups appear in the individual results sections.

\subsubsection{Artifact Availability}\label{sec:eval:artifact}
The source code for the fine-grained PyTorch memory allocation extension, the CXL-aware allocation policy is publicly available at \url{https://github.com/DandinPower/CXL-Aware-LLM-Finetuning}.

\begin{table}[h]
    \vspace{-0.1in}
    \centering
    \small
    \caption{Hardware and Software Specification for Experimental Setup.}
    \label{tab:system_spec}
    \begin{tabularx}{\columnwidth}{@{} >{\RaggedRight}X c l @{}}
        \toprule
        \textbf{Component} & \textbf{Specification} \\
        \midrule
        \multicolumn{2}{@{}l}{\textit{Hardware}} \\
        OS & Ubuntu 24.04 LTS \\
        Linux Kernel & v6.9 \\
        CPU & 1 $\times$ Intel(R) Xeon(R) 6780E \\
        GPUs & 2 $\times$ NVIDIA H100 80GB PCIe \\
        PCIe & PCIe 5.0 (x16 links for GPUs and AICs) \\
        Local DRAM & 512 GB (4 $\times$ 128 GB DDR5-6400) \\
        CXL AICs. (Config. A) & 1 $\times$ CXA-8F2W (512 GB AIC) \\
        CXL AICs. (Config. B) & 2 $\times$ CXA-4F1W (256 GB AIC) \\
        \midrule
        \multicolumn{2}{@{}l}{\textit{Software}} \\
        NUMA Control & numactl, libnuma 2.0.19 \\
        PyTorch & torch 2.5.1 \\
        Model & transformers 4.47.1 \\
        Framework & deepspeed 0.16.0 \\
        \bottomrule
    \end{tabularx}
    \vspace{-0.05in}
\end{table}

\begin{figure*}[!t]
    \centering
    \subfloat[7B model in a single-GPU scenario]{\includegraphics[height=1.28in]{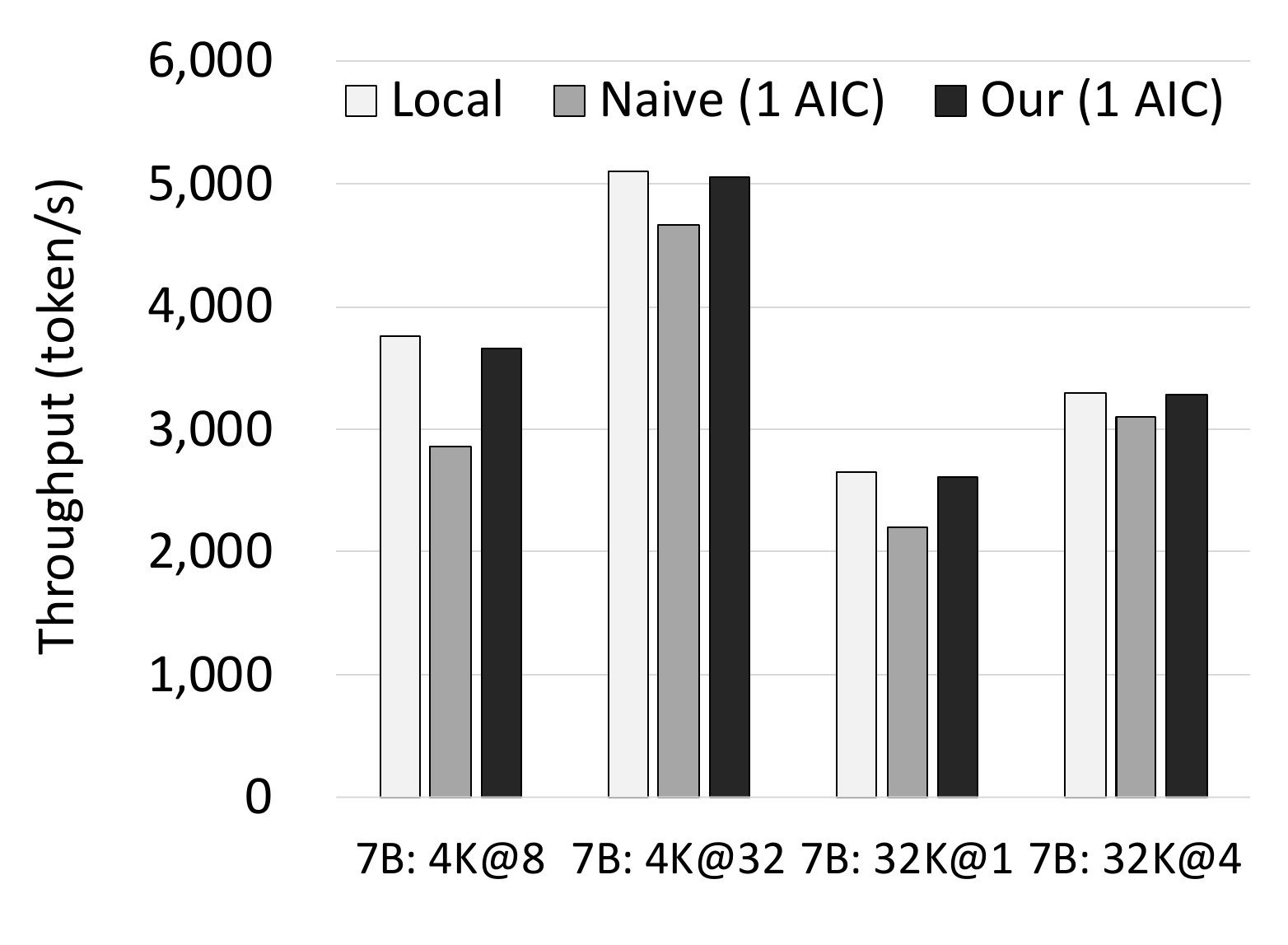}}
        \vspace{-0.01in}
    \subfloat[7B model in a dual-GPU scenario]{\includegraphics[height=1.28in]{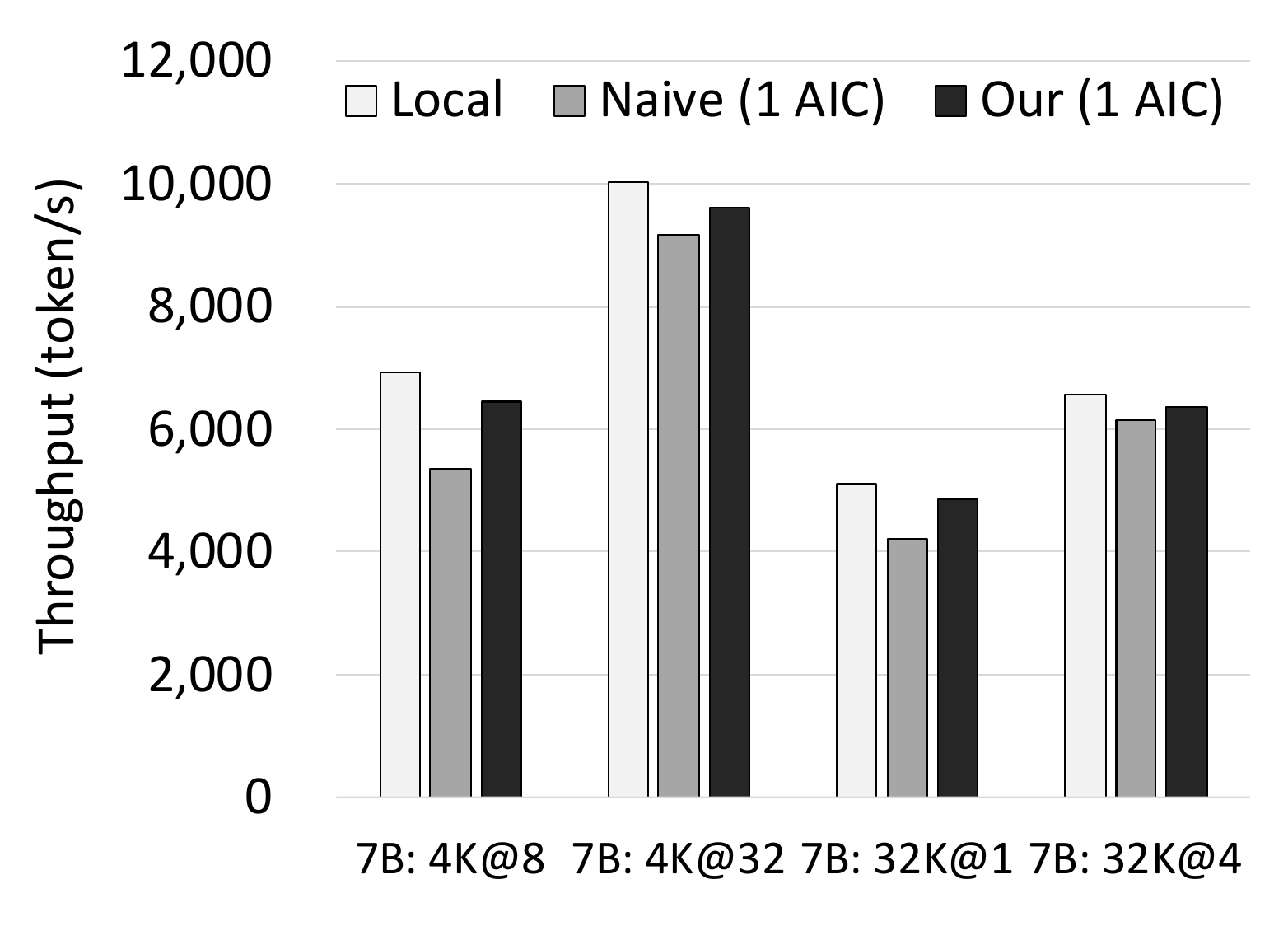}}
        \vspace{-0.01in}
    \subfloat[12B models in a single-GPU scenario]{\includegraphics[height=1.28in]{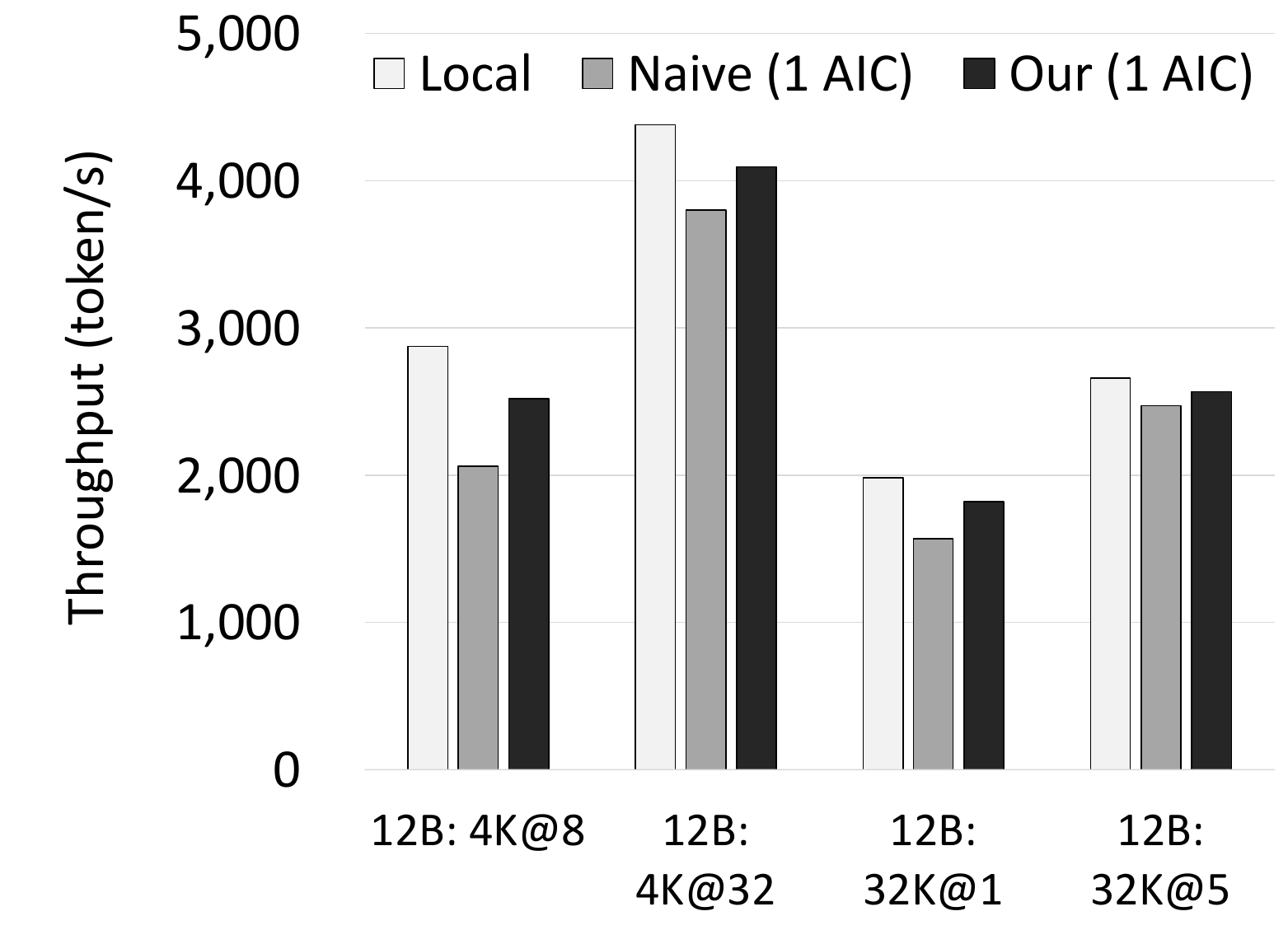}}
        \vspace{-0.01in}
    \subfloat[12B models in a dual-GPU scenario]{\includegraphics[height=1.28in]{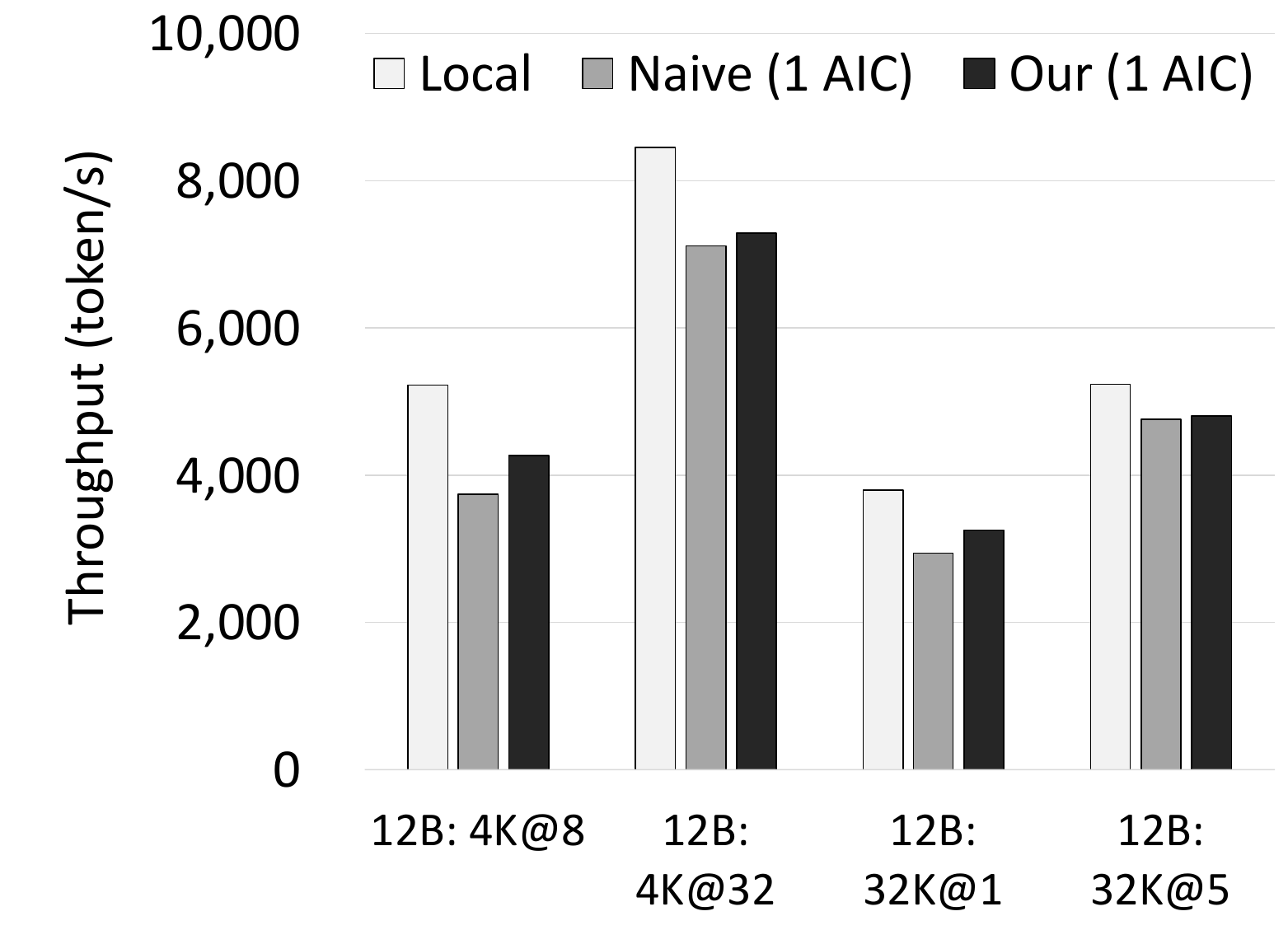}}
        \vspace{-0.01in}
    \caption{Training throughput comparison of three single-AIC configurations: baseline (local DRAM only), naive CXL (interleaving), and CXL-aware allocation (labeled as Our), evaluated across varying models, context lengths and batch sizes.}
    \vspace{-0.15in}
    \label{fig:eval_singleaic}
\end{figure*}

\begin{figure*}[!t]
    \centering
    \subfloat[7B model in a single-GPU scenario]{\includegraphics[height=1.28in]{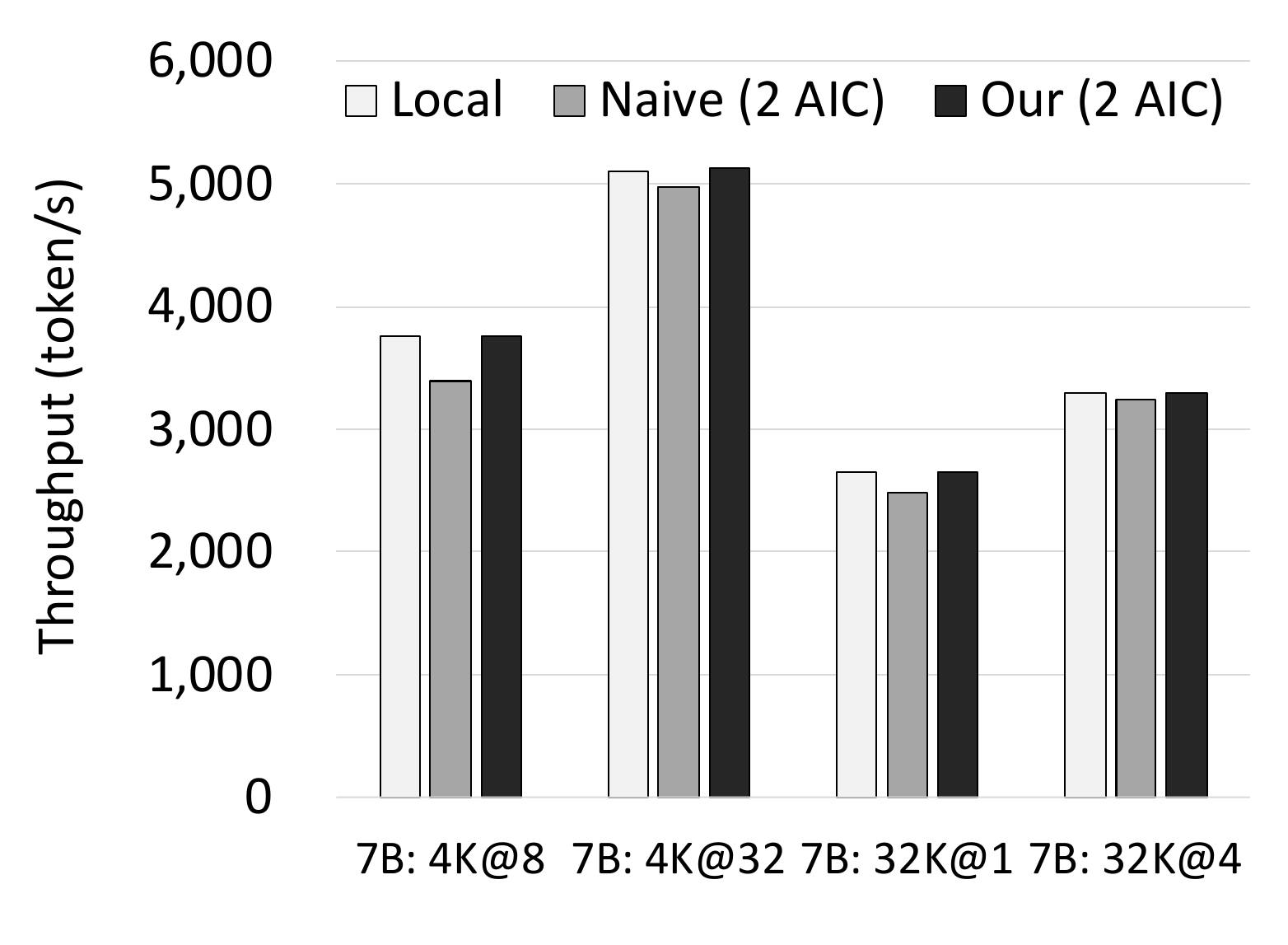}}
        \vspace{-0.01in}
    \subfloat[7B model in a dual-GPU scenario]{\includegraphics[height=1.28in]{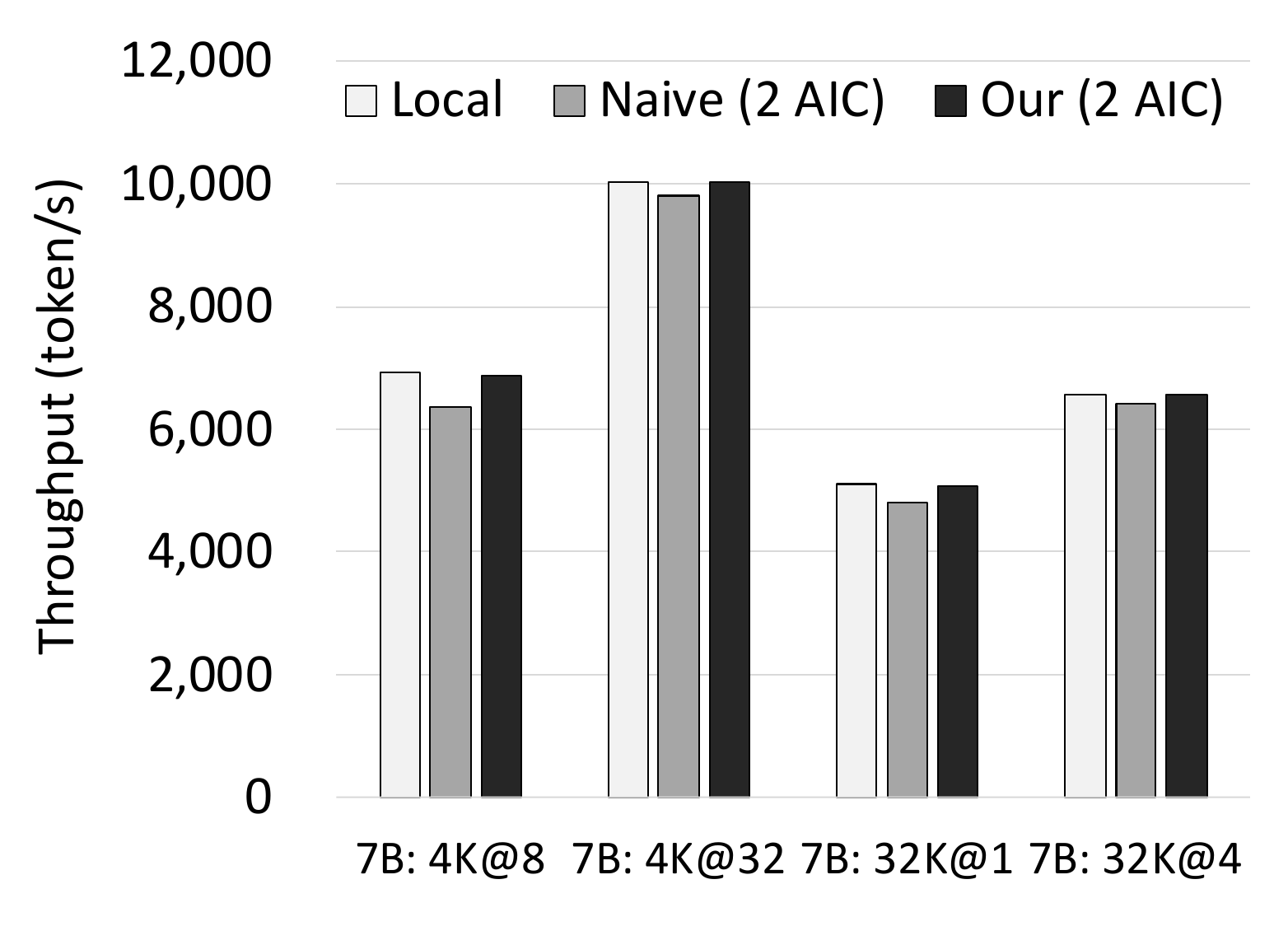}}
        \vspace{-0.01in}
    \subfloat[12B models in a single-GPU scenario]{\includegraphics[height=1.28in]{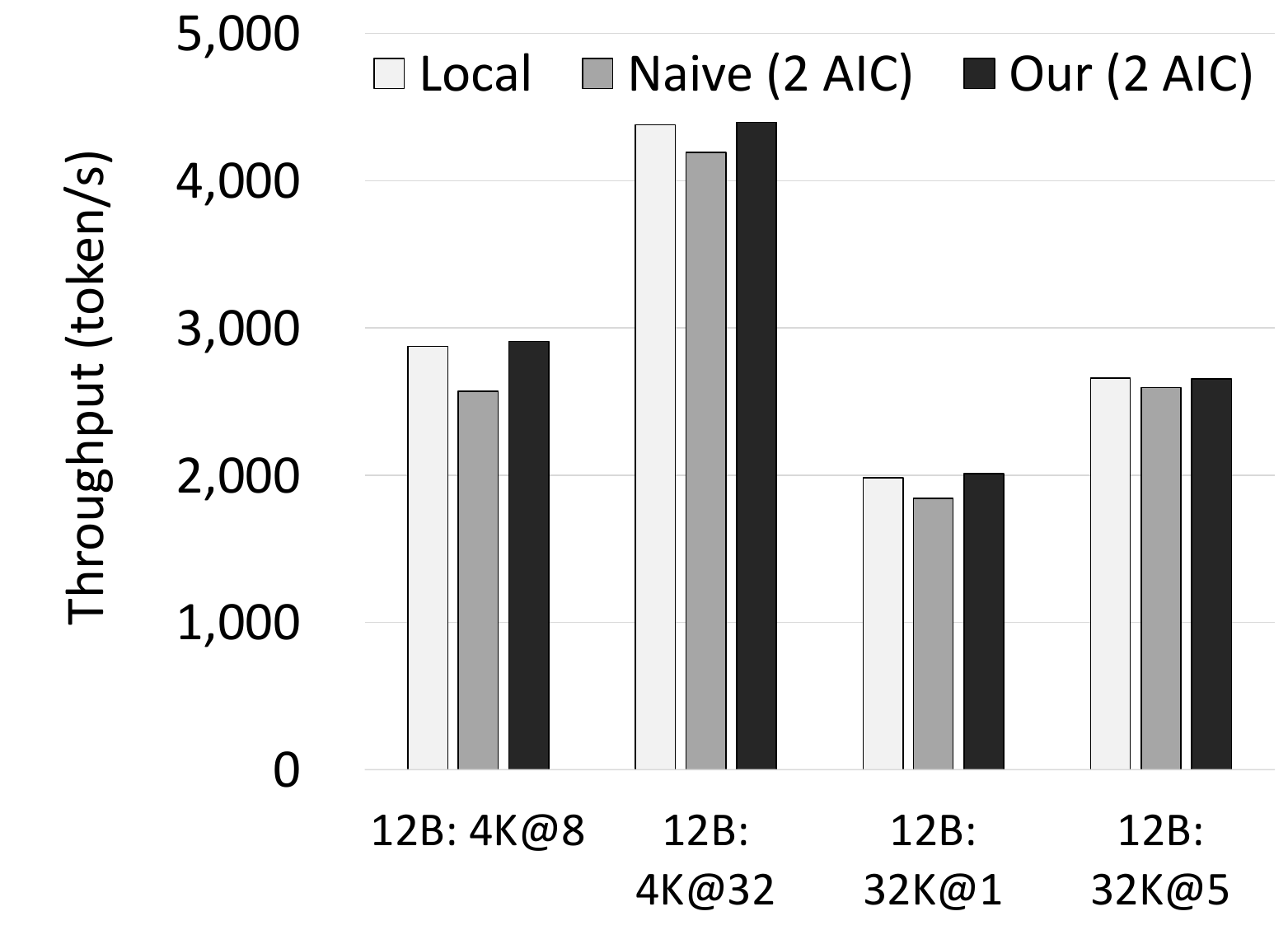}}
        \vspace{-0.01in}
    \subfloat[12B models in a dual-GPU scenario]{\includegraphics[height=1.28in]{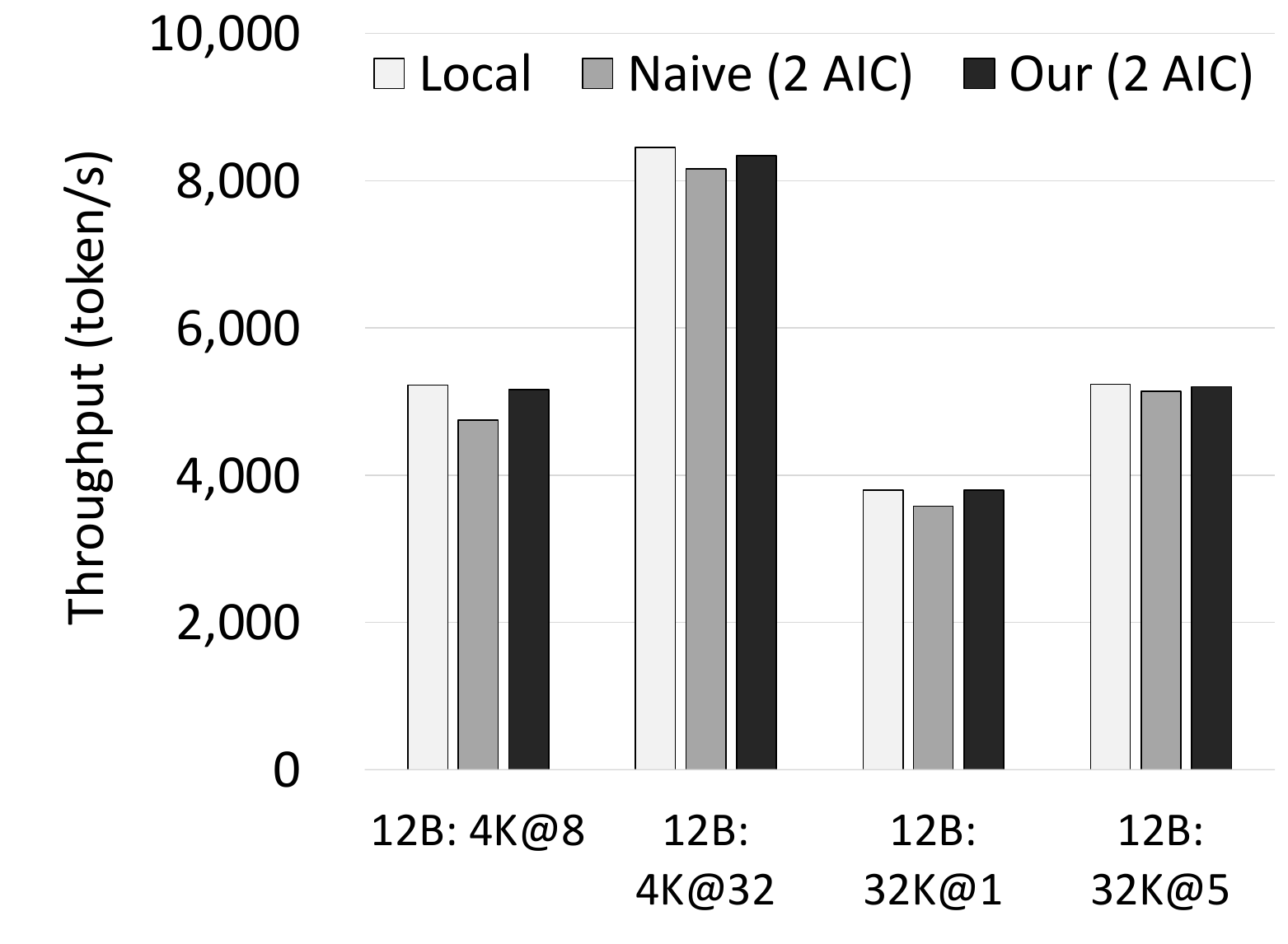}}
        \vspace{-0.01in}
    \caption{Training throughput comparison of three dual-AIC configurations: baseline (local DRAM only), naive CXL (interleaving), and CXL-aware allocation (labeled as Our), evaluated across varying models, context lengths and batch sizes.}
    \vspace{-0.15in}
    \label{fig:eval_dualaic}
\end{figure*}

\vspace{-0.1in}
\subsection{Performance Evaluation with a Single CXL AIC}\label{sec:eval:singleaic}
The single-AIC setup, corresponding to Config. A in Table~\ref{tab:system_spec}, serves as the baseline environment. Three configurations are evaluated to quantify the impact of CXL-attached memory. The first is the \textbf{baseline}, where all data reside in local DRAM. The second, \textbf{naive CXL}, combines 128 GiB of local DRAM with 512 GiB of CXL-attached memory using a \texttt{uniform numactl --interleave=all} policy. The third, \textbf{CXL-aware allocation}, employs the same capacities but applies the proposed fine-grained memory allocation extension and CXL-aware memory allocator.

Figure~\ref{fig:eval_singleaic}(a) shows single-GPU throughput for a 7B model under varying context lengths (4K--32K) and batch sizes (1–32). Relative to the baseline (normalized to 100\%), the naive CXL configuration sustains only 76\%–94\% throughput, depending on workload mix. Workloads dominated by FWD and BWD passes experience smaller losses because the latency-sensitive STEP occupies a smaller runtime fraction. By contrast, CXL-aware allocation restores throughput to 97\%--99\%, reducing the degradation to just 1\%–3\% compared to the local-DRAM baseline and outperforming naive CXL by up to 21\%.

Figure~\ref{fig:eval_singleaic}(b) presents dual-GPU results for the same 7B model. The naive CXL setup achieves 77\%–93\% of baseline throughput, while CXL-aware allocation improves this to 93\%–97\%, narrowing the gap to 3\%–7\%. The smaller gain compared with the single-GPU case stems from bandwidth contention: both GPUs share a single AIC, effectively halving available read/write bandwidth. This bandwidth constraint limits both FWD and BWD phases, underscoring the benefit of deploying multiple AICs. Even with identical total capacity, multiple AICs aggregate bandwidth more effectively, allowing each GPU to access sufficient bandwidth and thus recover full performance. Section~\ref{sec:eval:dualaic} further analyzes this multi-AIC configuration.

Figure~\ref{fig:eval_singleaic}(c) extends the analysis to a 12B model under a single-GPU setup. The naive CXL configuration reaches 72\%–93\% of baseline throughput, while CXL-aware allocation raises this to 88\%–96\%, reflecting a 4\%–12\% shortfall relative to the local-DRAM baseline but up to 16\% improvement over naive CXL. Although the benefit remains evident, the smaller margin arises from latency sensitivity rather than bandwidth limits. In this case, local DRAM cannot hold all latency-critical data, forcing some onto the slower AIC. This co-location increases access latency, which is alleviated when dual AICs are employed—providing additional bandwidth that effectively reduces latency. Section~\ref{sec:eval:dualaic} presents detailed results for this case, where CXL-aware allocation can even surpass the local-DRAM baseline.

Finally, Figure~\ref{fig:eval_singleaic}(d) reports dual-GPU throughput for the 12B model with one AIC. The naive CXL setup sustains 72\%–91\% of baseline performance, while CXL-aware allocation increases it to 82\%–92\%, offering up to 10\% improvement over naive CXL. Overall, across all single-AIC setups, CXL-aware allocation offers an average improvement of 11.40\% over the naive CXL policy. Nonetheless, concurrent bandwidth contention and latency sensitivity constrain further gains. The following section evaluates how CXL-aware allocation alleviates these bottlenecks under a dual-AIC environment.

\begin{figure*}[!t]
    \centering
    \subfloat[7B model in a single-GPU scenario]{\includegraphics[height=1.2in]{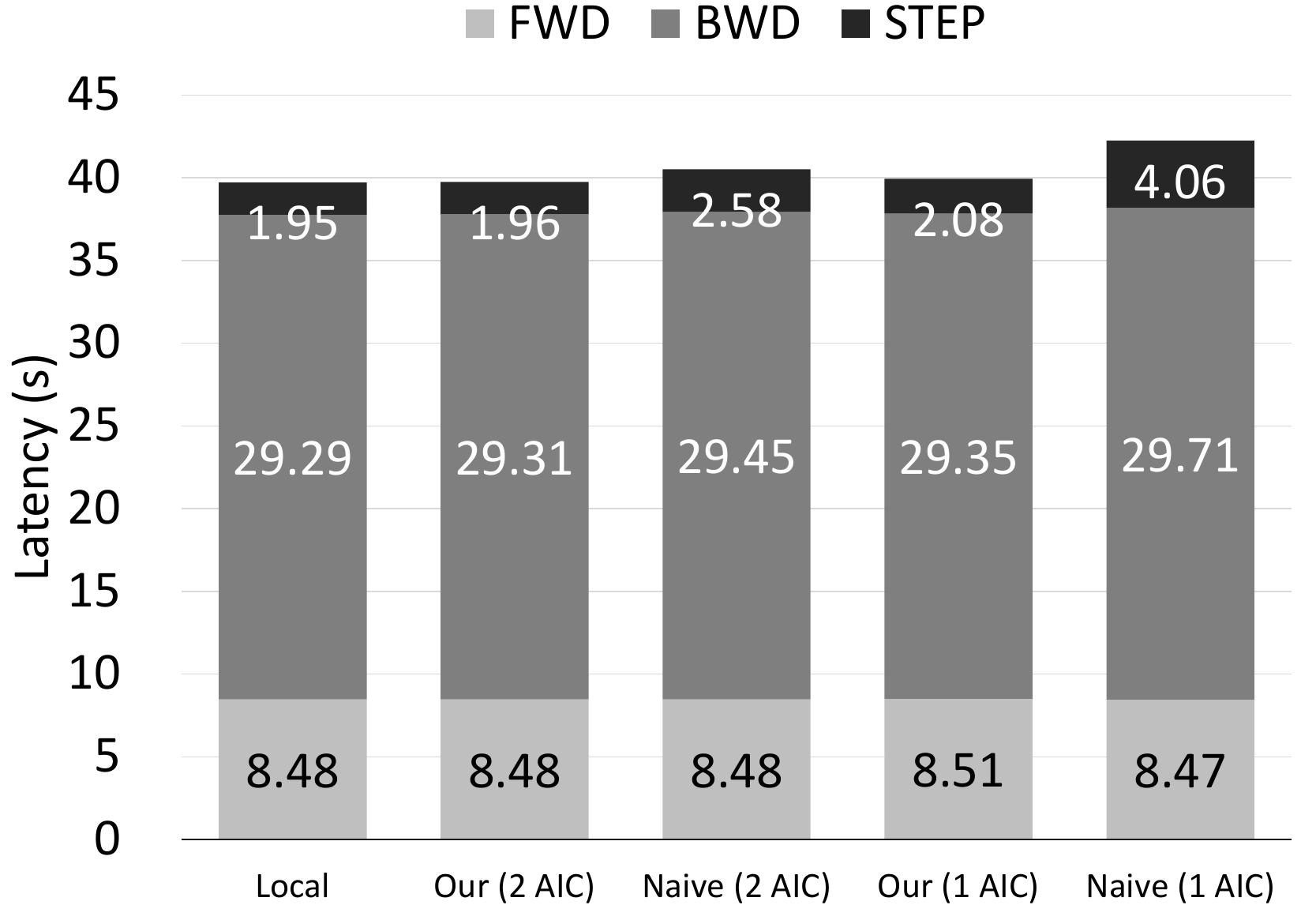}}
        \vspace{-0.01in}
    \subfloat[7B model in a dual-GPU scenario]{\includegraphics[height=1.2in]{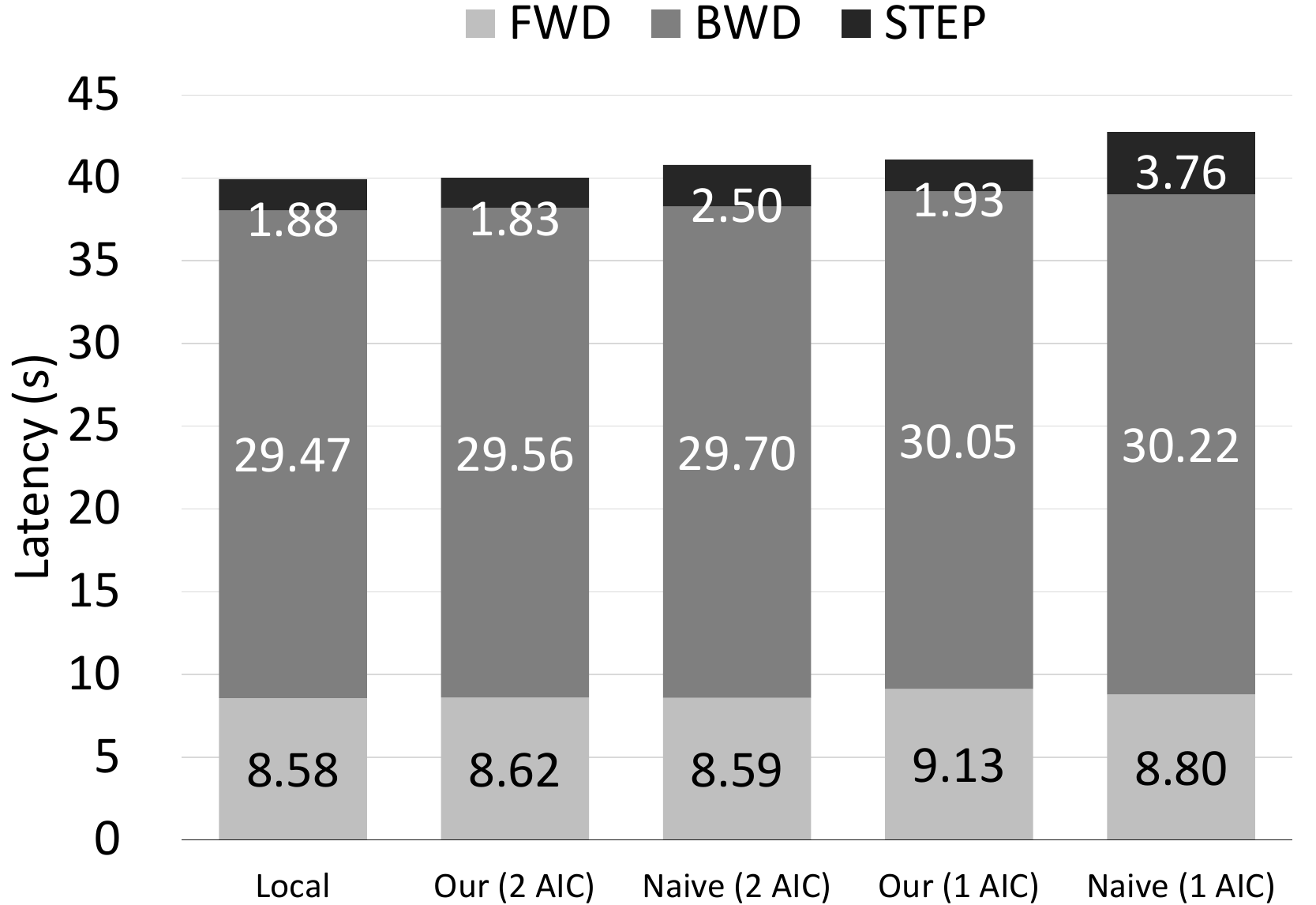}}
        \vspace{-0.01in}
    \subfloat[12B models in a single-GPU scenario]{\includegraphics[height=1.2in]{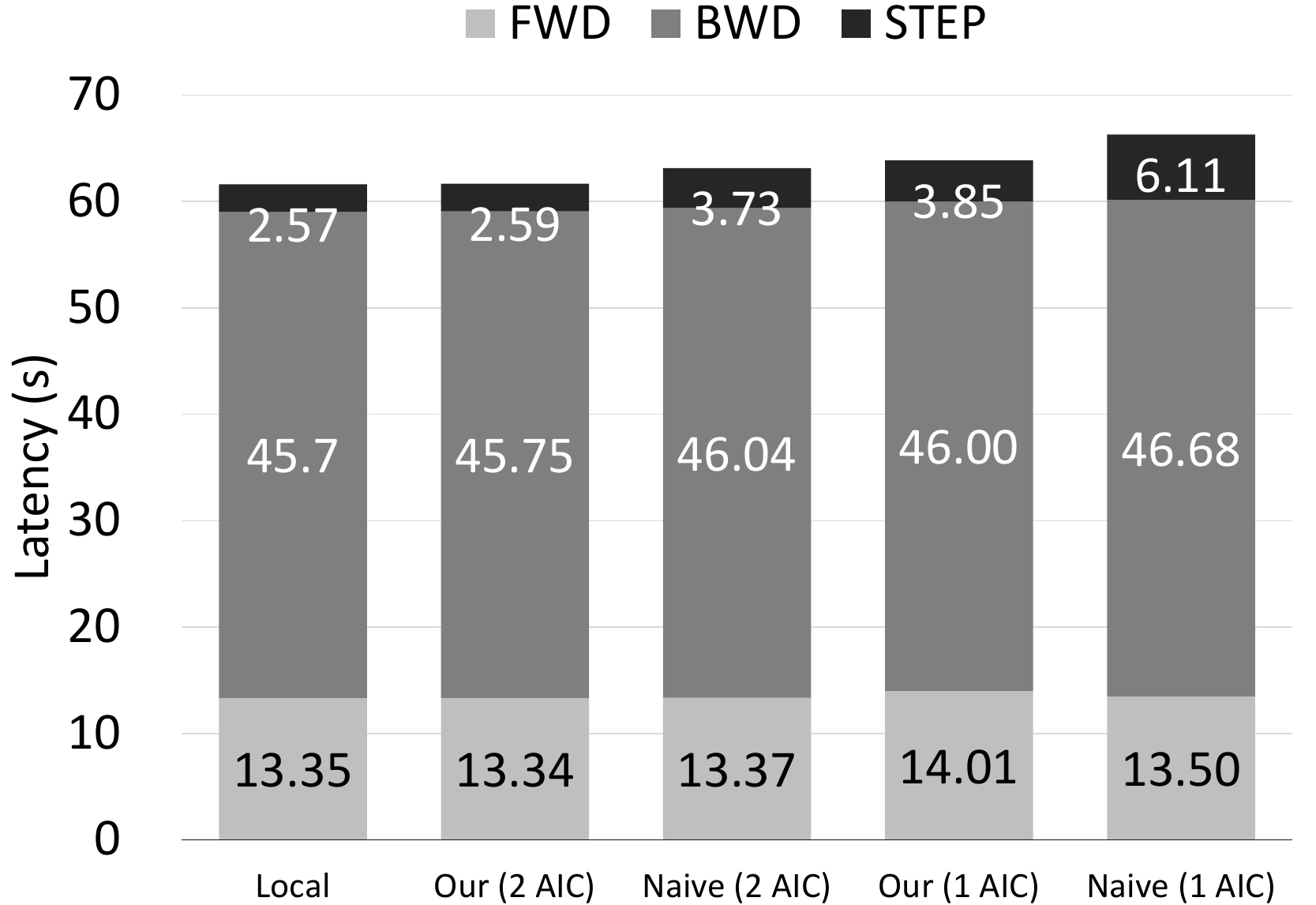}}
        \vspace{-0.01in}
    \subfloat[12B models in a dual-GPU scenario]{\includegraphics[height=1.2in]{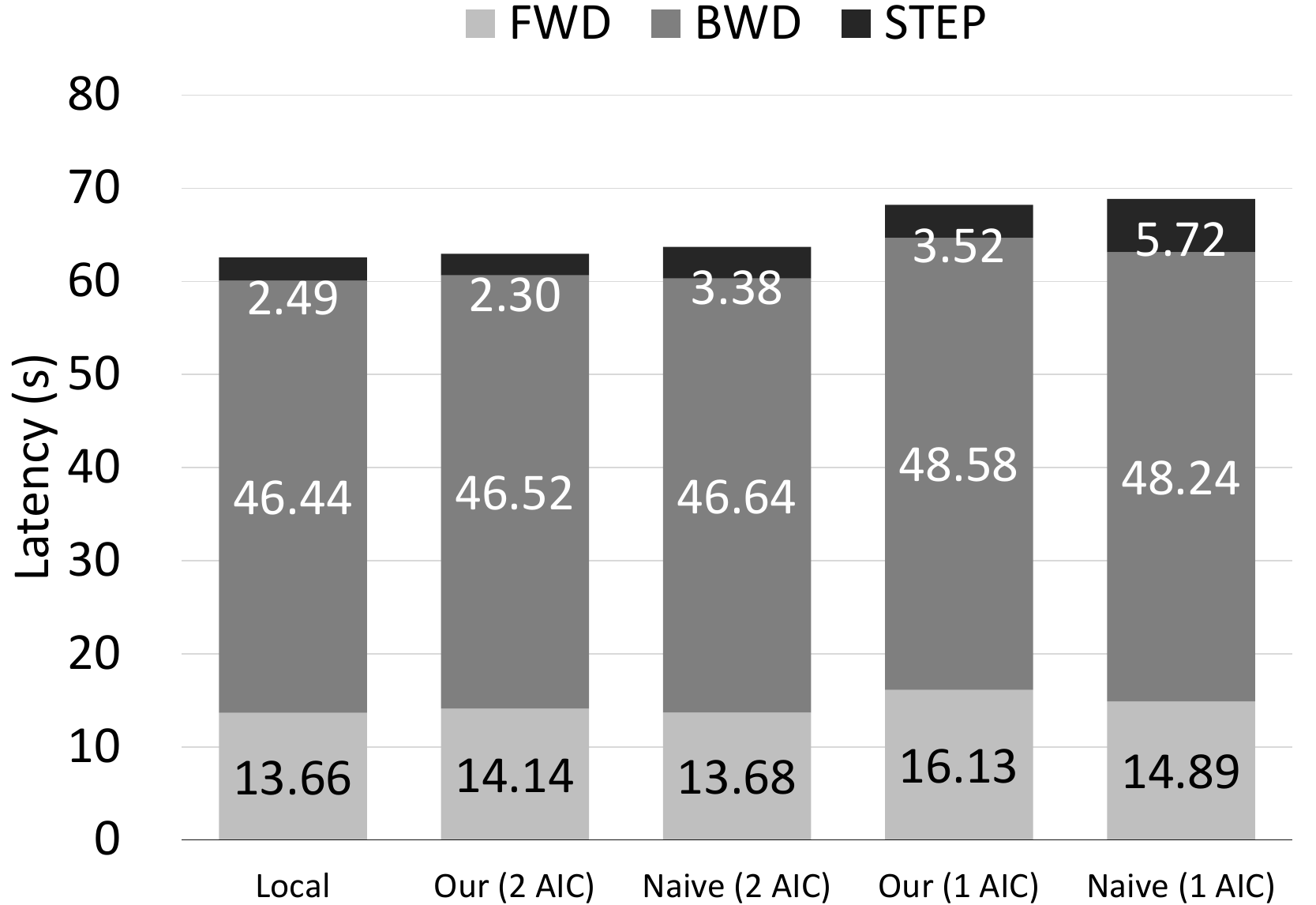}}
        \vspace{-0.01in}
    \caption{Training latency decomposition for a long-context workload (32K context, batch size 4), showing that CXL-aware allocation consistently outperforms the naive CXL policy and nearly matches the local-DRAM baseline performance on both single- and dual-AIC systems.}
    \vspace{-0.15in}
    \label{fig:eval_detail}
\end{figure*}

\vspace{-0.1in}
\subsection{Performance Evaluation with Dual CXL AICs}\label{sec:eval:dualaic}
The evaluation next turns to the dual-AIC scenario, designated as Config. B in Table~\ref{tab:system_spec}. Three configurations are used to establish the impact of CXL-attached memory. The first is the \textbf{baseline}, where all data remain in local DRAM. The second is \textbf{naive CXL}, which pairs 128 GiB of DRAM with two 256 GiB AICs under a naive \texttt{numactl --interleave=all} policy. The third is \textbf{CXL-aware allocation}, which uses the same capacities but applies the proposed algorithm.

Figure~\ref{fig:eval_dualaic}(a) presents single-GPU throughput for a 7B model across the same range of context lengths and batch sizes used earlier in the single-AIC configuration. While the naive CXL policy results in a 2\% to 9\% performance drop compared to the baseline, the proposed CXL-aware allocation restores performance to 100\% of the baseline, showing that no performance is lost when CXL-attached memory is managed intelligently. The algorithm achieves this by automatically interleaving latency-tolerant data across both AICs, thereby capitalizing on their aggregate bandwidth. These results demonstrate that with workload-aware allocation, a dual-card CXL configuration can fully match native DRAM performance.

Figure~\ref{fig:eval_dualaic}(b) shows dual-GPU throughput for a 7B model across the same context lengths and batch sizes used in the single-AIC setup. The naive CXL policy lowers performance by 2\% to 8\% relative to the baseline, whereas CXL-aware allocation restricts the loss to no more than 1\%. In the single-AIC case shown in Figure~\ref{fig:eval_singleaic}(b), bandwidth contention remained a limiting factor even with CXL-aware allocation. The dual-AIC setup alleviates this issue, allowing CXL-aware allocation to exploit the combined resources of both cards and largely eliminate the penalty, while naive CXL continues to falter due to poor placement.

Figure~\ref{fig:eval_dualaic}(c) illustrates single-GPU throughput for a 12B model across the same range of context lengths and batch sizes. The naive CXL policy reduces throughput by 2\% to 11\% compared to the baseline, whereas the CXL-aware allocation restores performance to 100\%–101\% of the baseline. In the single-AIC case (Figure~\ref{fig:eval_singleaic}(c)), limited local DRAM capacity constrained performance even with CXL-aware placement. That limitation still exists in the dual-AIC setup, but the CXL-aware allocation mitigates its impact by leveraging the combined bandwidth of both cards alongside local DRAM. The latency-first greedy algorithm places latency-sensitive data in local DRAM whenever possible and assigns the remaining data across local DRAM and both AICs to aggregate bandwidth and minimize overall memory-access latency. The naive CXL policy, by contrast, continues to underperform due to unoptimized placement.

Finally, Figure~\ref{fig:eval_dualaic}(d) reports dual-GPU throughput for the 12B model. The naive CXL policy reduces throughput by 2\% to 9\% relative to the baseline, while the CXL-aware allocation limits the loss to at most 1\%. Overall, across all dual-AIC setups, CXL-aware allocation offers an average improvement of 5.56\% over the naive CXL policy. In the single-AIC case (Figure~\ref{fig:eval_singleaic}(d)), bandwidth contention and limited DRAM capacity remained bottlenecks even with CXL-aware allocation. The dual-AIC configuration resolves these issues: CXL-aware allocation places latency-sensitive data appropriately and distributes remaining allocations across all CXL NUMA nodes to fully aggregate bandwidth. This automated interleaving enables both CPU and GPU workloads to exploit the combined bandwidth of the dual cards. The resulting performance, effectively matching the local-DRAM baseline, underscores the importance of CXL-aware allocation for intelligently orchestrating underlying hardware resources.

\vspace{-0.15in}
\subsection{Latency Decomposition}\label{sec:eval:latency_decomposition}
While the previous section established end-to-end performance, this section breaks down the latency of FWD, BWD, and STEP for each configuration to further analyze latency and the impact of CXL-aware allocation. The system setup and configurations follow those used earlier: the single-AIC scenario corresponds to Config. A in Table~\ref{tab:system_spec}, and the dual-AIC scenario corresponds to Config. B. Five configurations are compared in the same order as shown in Figure~\ref{fig:eval_detail}. The first is the baseline, where all data remain in local DRAM. The second is CXL-aware allocation (2 AIC), which pairs 128 GiB of DRAM with two 256 GiB AICs and applies the proposed extension and allocator. The third is naive CXL (2 AIC), which uses the same capacities under a naive \texttt{numactl --interleave=all} policy. The fourth is CXL-aware allocation (1 AIC), which combines 128 GiB of local DRAM with 512 GiB of CXL-attached memory and applies the proposed CXL-aware memory allocator. The fifth is naive CXL (1 AIC), which uses the same capacities under the same interleave-all policy.

Figure~\ref{fig:eval_detail} presents a detailed latency breakdown of the training process into FWD, BWD, and STEP phases, revealing how memory configurations affect each stage. In the 7B single-GPU scenario shown in Figure~\ref{fig:eval_detail}(a), FWD and BWD remain around $8.5 s$ and $29.3 s$ across all policies. The major separation arises in STEP: the baseline completes in $\approx1.95 s$, naive CXL (1 AIC) inflates it to $\approx4.06 s$, and naive CXL (2 AIC) reduces it slightly to $\approx2.58 s$. In contrast, CXL-aware allocation lowers STEP to $\approx2.08 s$ on one AIC and $\approx1.96 s$ on two AICs, essentially matching the baseline. This is because the naive interleave policy places latency-critical optimizer states on AIC memory, forcing frequent CPU accesses through a high-latency path. On the other hand, the CXL-aware allocation pins latency-sensitive tensors in local DRAM and pushes only bandwidth-tolerant data to CXL, eliminating this penalty.

Figure~\ref{fig:eval_detail}(b) shows the 7B dual-GPU scenario, where bandwidth contention on the CXL interconnect becomes more pronounced. With a single AIC, both GPUs compete for bandwidth, slowing FWD and BWD even under CXL-aware allocation (1 AIC). This bottleneck highlights the limitation of a single CXL device in a multi-GPU environment. However, CXL-aware allocation (2 AIC) resolves this issue by intelligently distributing memory accesses across both AICs, providing sufficient aggregate bandwidth and restoring performance to within 0.2\% of the local-DRAM baseline. For the larger 12B model in a single-GPU setup (Figure~\ref{fig:eval_detail}(c)), FWD and BWD remain similar across configurations, but STEP dominates. The baseline STEP is $\approx 2.57 s$; naive CXL (1 AIC) raises it to $\approx 6.11 s$, while CXL-aware allocation (1 AIC) lowers it to $\approx 3.85 s$. CXL-aware allocation (2 AIC) further cuts STEP to $\approx 2.59 s$, effectively equal to the baseline. The small residual gap in the single-AIC aware case aligns with earlier findings: local DRAM cannot accommodate all latency-critical tensors for 12B, so part of STEP still touches AIC memory. With only one card, this additional latency remains unavoidable.

Finally, the most demanding configuration, the 12B model with dual GPUs, shown in Figure~\ref{fig:eval_detail}(d), demonstrates the combined effects of latency sensitivity and bandwidth contention. Single-AIC configurations suffer substantial degradation, with total latency increasing by up to 10\% over the baseline because the AIC is fully saturated. Both FWD and BWD phases, as well as STEP, slow down. In contrast, CXL-aware allocation (2 AIC) effectively manages the hardware resources: by distributing allocations across local DRAM and both CXL NUMA nodes, it aggregates bandwidth while prioritizing latency-sensitive data, eliminating CXL-induced overhead and matching the performance of the local-DRAM baseline.

\vspace{-0.15in}
\subsection{Scaling to Multi-GPU and Multi-Node Systems}
\label{sec:eval:scalability_implications}

The insights from Section~\ref{sec:eval:dualaic} and Section~\ref{sec:eval:latency_decomposition} naturally extend to systems scaling beyond the evaluated two-GPU baseline. For multi-GPU nodes, scalability hinges on maintaining a 1:1 ratio between effective CXL bandwidth and GPU PCIe bandwidth. When this bandwidth parity is met, either via dedicated CXL AICs or sufficient aggregate bandwidth, four- and eight-GPU nodes will follow the same offloading trends. Latency-sensitive optimizer data remains securely in local DRAM, while latency-tolerant data is striped across CXL devices to eliminate contention and approach DRAM-level performance. If CXL bandwidth is bottlenecked by too few devices sharing too many GPUs, CXL-aware allocation still avoids the severe latency penalties of naive CXL, though FWD/BWD transfers become bandwidth-bound and maximum throughput is reduced. Furthermore, extending to multi-node training does not change the core offloading dynamics. While the data parallelism of DeepSpeed ZeRO requires cross-node GPU communication for distributed collectives, CPU-offloaded tensors are still partitioned and accessed locally by the CPU and GPU residing on the same node. Consequently, multi-node performance scales as a direct function of single-node efficiency. As long as the per-GPU CXL bandwidth is sufficient, CXL-aware allocation will effectively bridge the gap to the local DRAM baseline and consistently outperform naive CXL across distributed environments.

\begin{figure}[t]
    \centering
    \begin{minipage}{0.24\textwidth}
        \centering
        \includegraphics[width=1.75in]{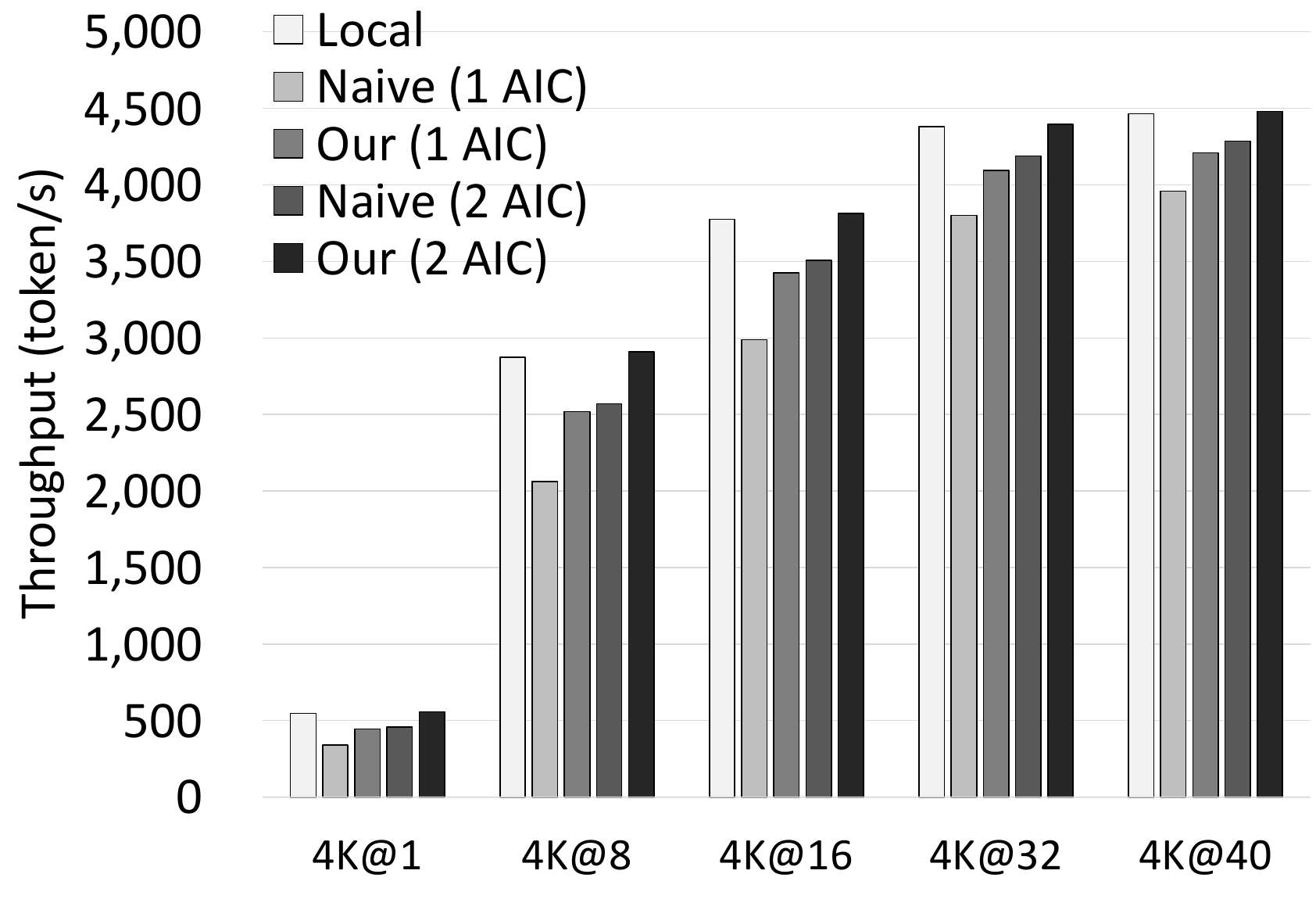}
        \vspace{-0.25in}
        \caption{Training throughput comparison across batch sizes for the 12B model with 4K context length.}
        \label{fig:batch_ablation_bar}
    \end{minipage}\hfill
    \begin{minipage}{0.24\textwidth}
        \centering
        \includegraphics[width=1.75in]{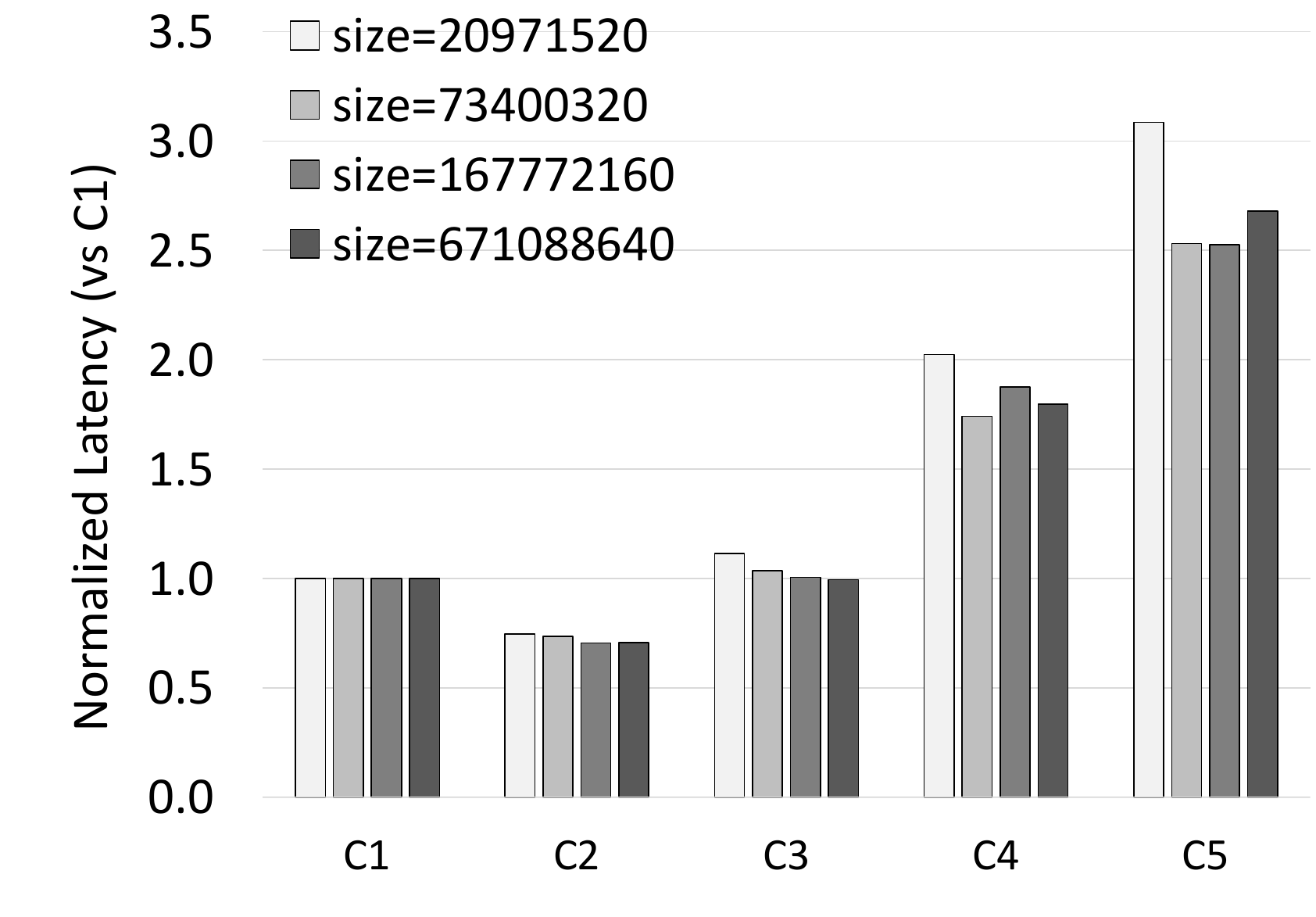}
        \vspace{-0.25in}
        \caption{CPU-based Adam optimizer step latency under different data-placement configurations.}
        \label{fig:cpu_optimizer_bench_extend}
    \end{minipage}
    \vspace{-0.05in}
\end{figure}

\vspace{-0.15in}
\subsection{Batch-Size Sensitivity Analysis}\label{sec:eval:batch_ablation}

Figure~\ref{fig:batch_ablation_bar} evaluates whether increasing batch size amortizes the latency overhead of CXL-attached memory. The experiment uses the 12B model with a 4K context length on one H100 GPU and varies the batch size from 1 to 40. The same memory-placement policies used in the previous evaluations are compared. Across all batch sizes, CXL-aware allocation consistently outperforms naive CXL. With one AIC, the improvement over naive CXL ranges from 30.8\% at BS=1 to 6.3\% at BS=40. With two AICs, the improvement similarly ranges from 21.7\% to 4.6\%. This trend occurs because larger batches increase the fraction of time spent in FWD and BWD, which are dominated by GPU compute time, thereby amortizing the impact of system memory access latency. However, the gap remains apparent even at BS=40: CXL-aware allocation reaches 4209.12 tokens/s compared with 3959.81 tokens/s for naive CXL in the 1-AIC case, and 4480.78 tokens/s compared with 4285.12 tokens/s in the 2-AIC case. These results show that increasing batch size reduces, but does not eliminate, the benefit of separating latency-sensitive optimizer data from latency-tolerant offloaded tensors.

\vspace{-0.15in}
\subsection{Cross-Platform and Cross-Workload Generalization}\label{sec:eval:ablation_on_5090}

\begin{table}[b]
    \centering
    \small
    \caption{Hardware Specification for Workstation Platform.}
    \label{tab:ablation_system_spec}
    \begin{tabularx}{\columnwidth}{@{} >{\RaggedRight}X c l @{}}
        \toprule
        \textbf{Component} & \textbf{Specification} \\
        \midrule
        OS & Ubuntu 24.04 LTS \\
        Linux Kernel & v6.9 \\
        CPU & 1 $\times$ Intel(R) Xeon(R) 6530P \\
        GPU & 1 $\times$ NVIDIA 5090 32GB PCIe \\
        PCIe & PCIe 5.0 (x16 links for GPU and AIC) \\
        Local DRAM & 256 GB (4 $\times$ 64 GB DDR5-4800) \\
        CXL AIC & 1 $\times$ R5X4 (256 GB AIC) \\
        \bottomrule
    \end{tabularx}
    \vspace{-0.05in}
\end{table}

\begin{figure*}[!t]
    \centering
    \subfloat[7B model in a 5090 platform]{\includegraphics[height=1.23in]{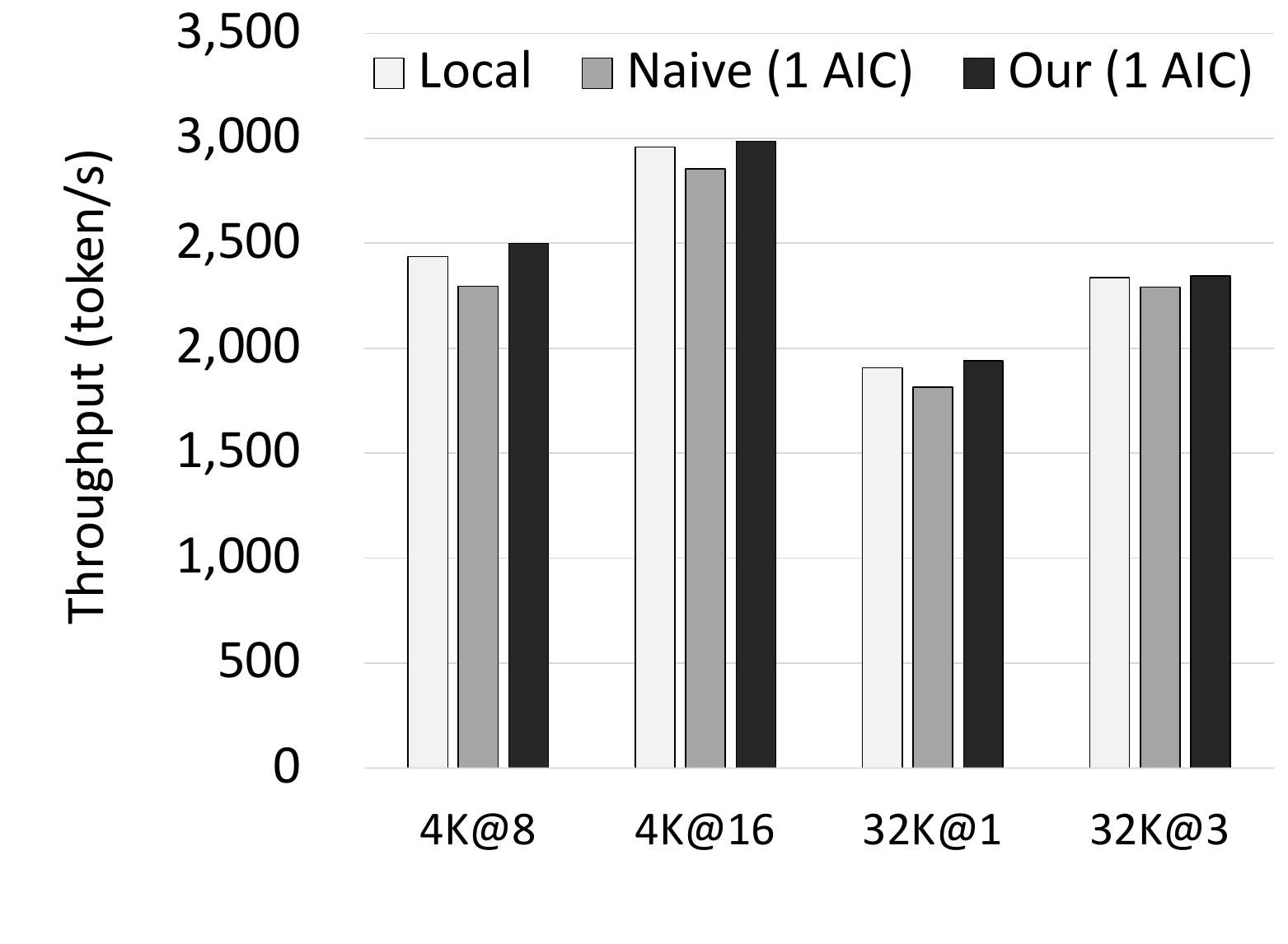}}
        \vspace{-0.01in}
    \subfloat[7B model in a Lion Optimizer]{\includegraphics[height=1.23in]{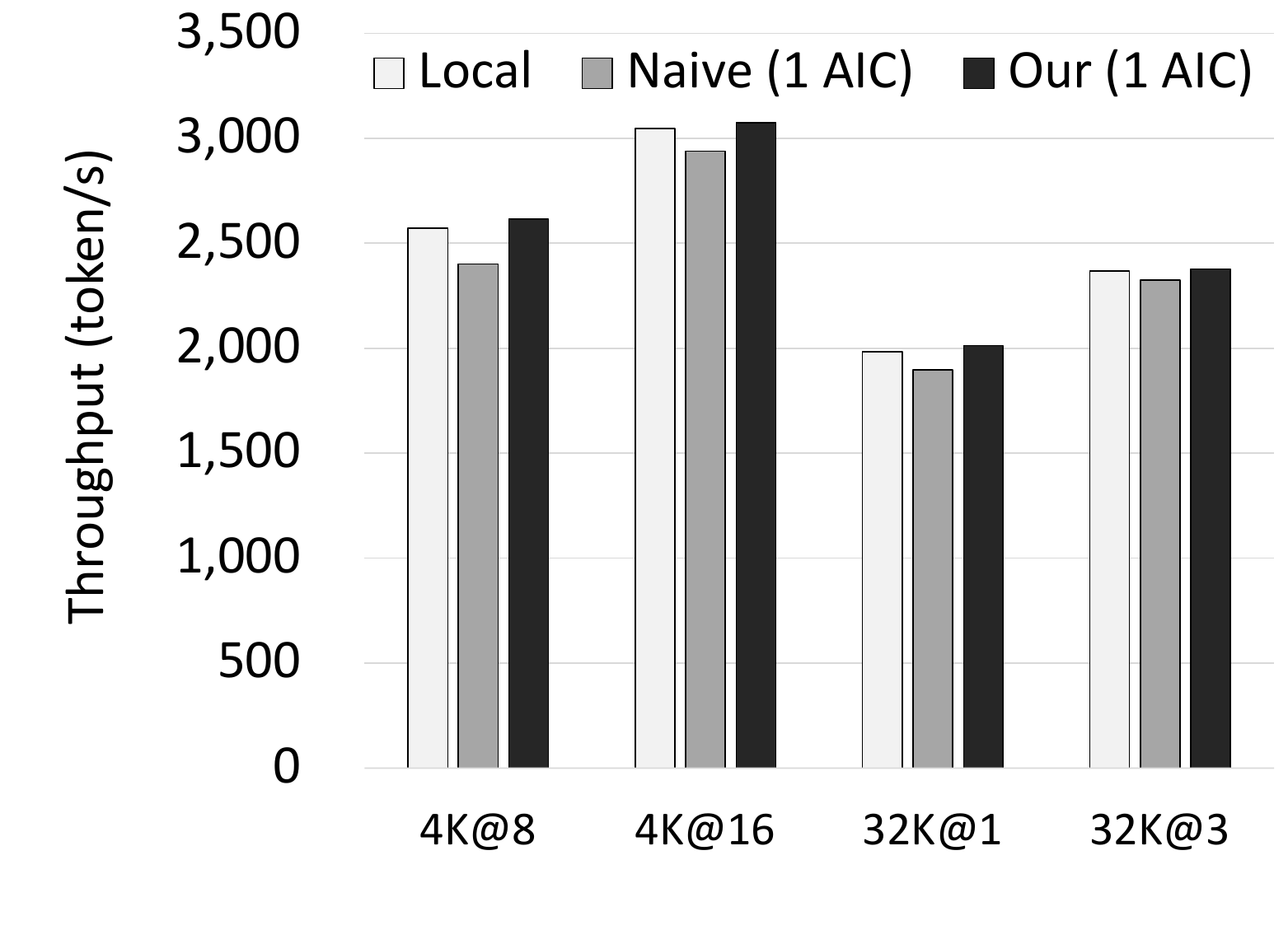}}
        \vspace{-0.01in}
    \subfloat[8B-A1B MoE Architecture]{\includegraphics[height=1.23in]{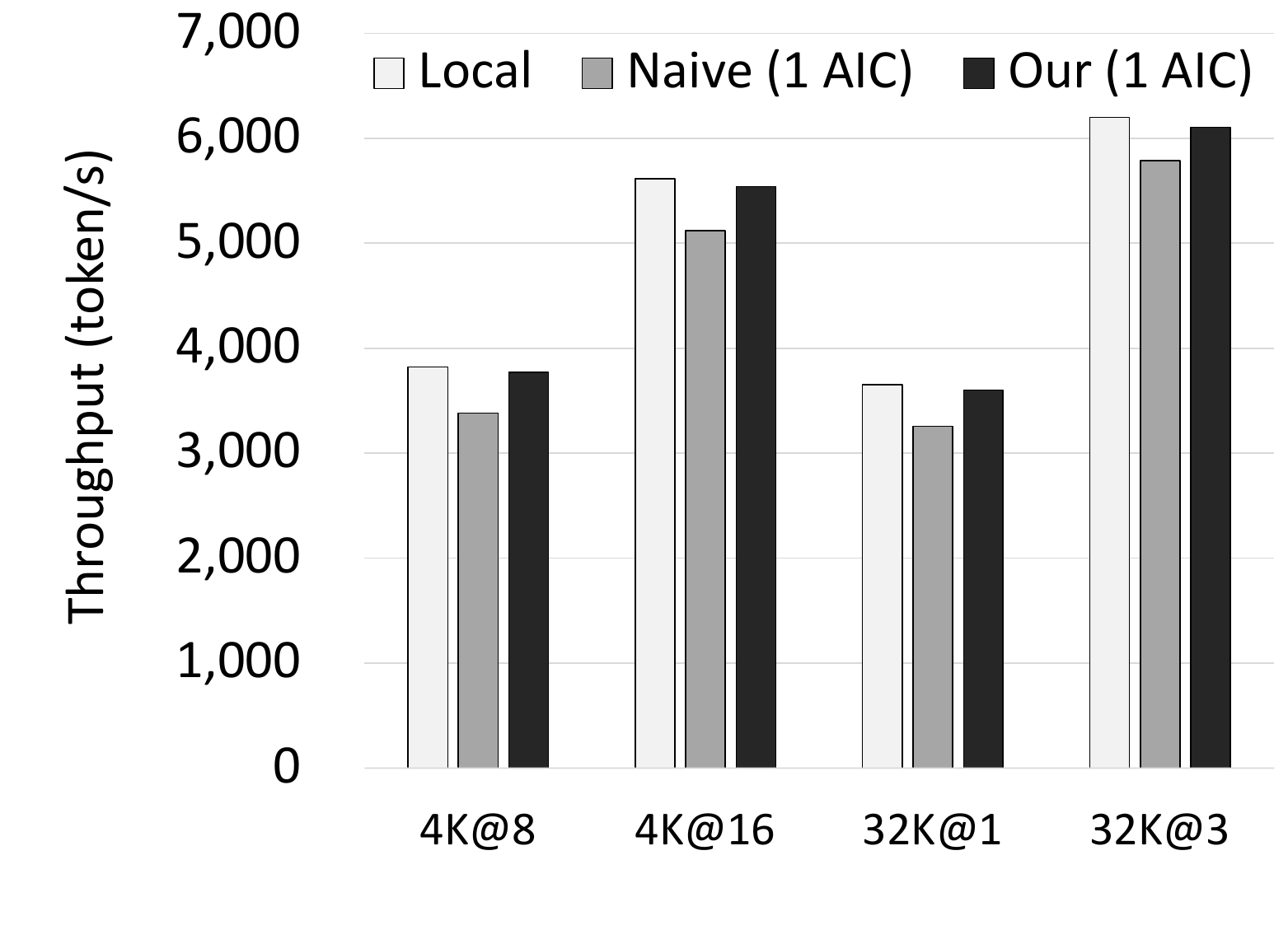}}
        \vspace{-0.01in}
    \subfloat[4B-4B Encoder-Decoder Architecture]{\includegraphics[height=1.23in]{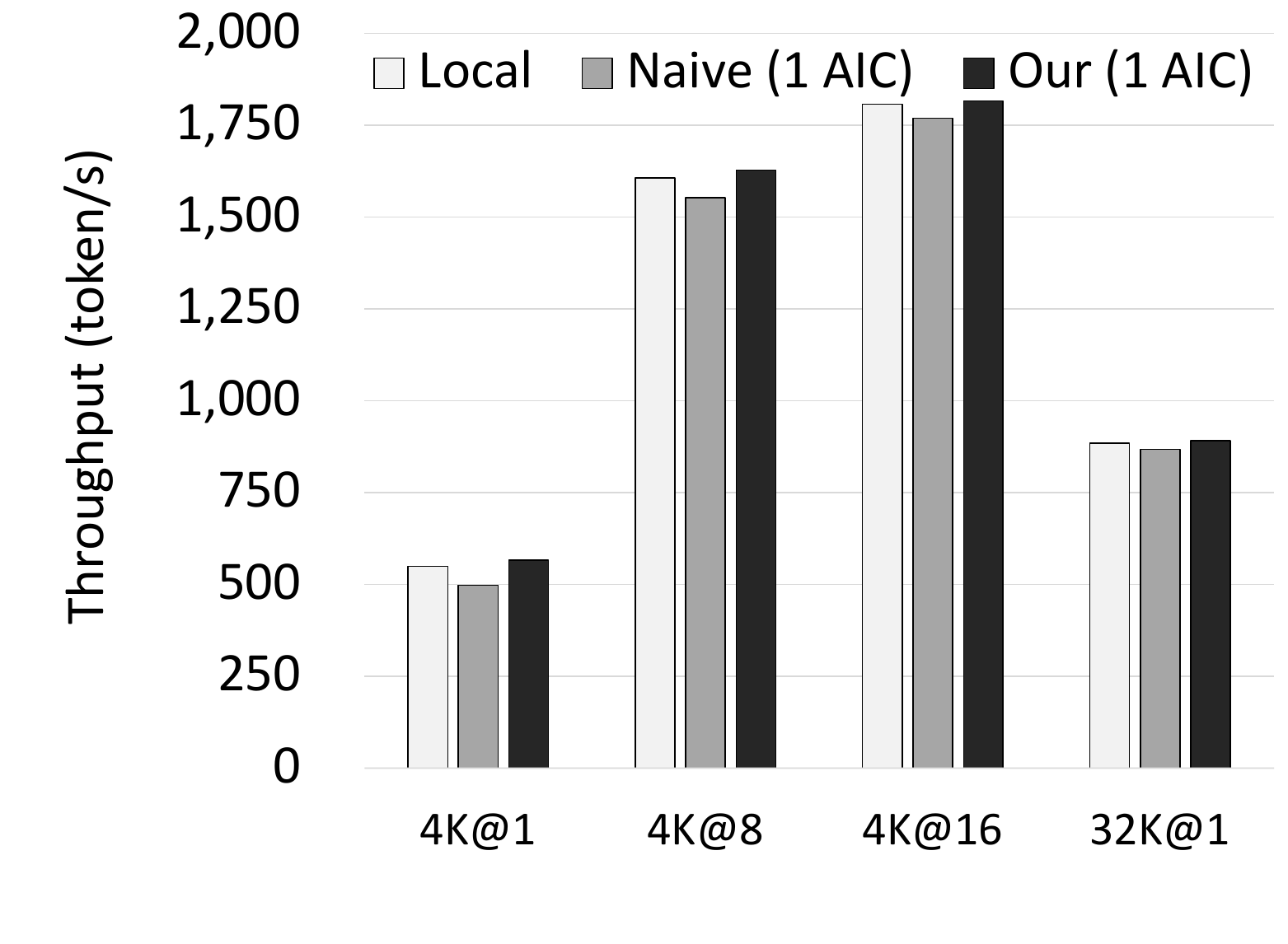}}
        \vspace{-0.01in}
    \caption{Training throughput comparison of three configurations: baseline (local DRAM only), naive CXL (interleaving), and CXL-aware allocation.}
    \vspace{-0.15in}
    \label{fig:ablation_studies}
\end{figure*}

To evaluate whether the proposed placement policy generalizes beyond the primary server-class platform, this section conducts cross-platform and cross-workload ablations on the workstation-class system detailed in Table~\ref{tab:ablation_system_spec}. Compared with the main setup in Table~\ref{tab:system_spec}, this platform features one NVIDIA 5090 GPU, 256~GB of local DRAM, and one 256~GB CXL AIC. This configuration is highly representative of workstation deployments, where commodity GPUs provide massive compute throughput but smaller VRAM capacities than data-center GPUs, and system-memory expansion is restricted by DIMM slots and CPU/chipset limitations. All ablations on this platform compare the same three policies: \textbf{baseline}, where the full 256~GiB of data resides in local DRAM; \textbf{naive CXL}, where local DRAM is constrained to 128~GiB and combined with 256~GiB of CXL-attached memory via hardware interleaving; and \textbf{CXL-aware allocation}, which uses the identical 128~GiB DRAM constraint but places tensors using the proposed latency-first policy.

Figure~\ref{fig:ablation_studies}(a) reports the Qwen2.5-7B performance on the 5090 platform across the plotted 4K and 32K settings. While naive CXL sustains 94\%--98\% of the local-DRAM throughput, CXL-aware allocation reaches 100\%--102\%, outperforming naive CXL by up to 8\% and confirming that the proposed policy preserves DRAM-like throughput on workstation architectures. This slightly above-baseline throughput is further analyzed using the CPU optimizer microbenchmark in Figure~\ref{fig:cpu_optimizer_bench_extend}. Configuration C1 places all optimizer data in local DRAM, C5 places all data in CXL-attached memory, and C2--C4 place 25\%, 50\%, and 75\% of optimizer data in CXL-attached memory, respectively. As the CXL fraction increases, optimizer-step latency generally rises because more CPU accesses must traverse the higher-latency CXL path. However, when only a minor fraction of optimizer data is placed in CXL, the mixed placement actually outperforms the local-DRAM baseline by exploiting aggregate bandwidth while retaining the most latency-sensitive accesses in local DRAM.

Beyond the primary hardware setup, the latency-aware placement principle remains robust across different optimizers and model architectures. Figure~\ref{fig:ablation_studies}(b) examines the 7B workload using the Lion optimizer, which maintains only a single momentum state rather than both momentum and variance states~\cite{chen2023symbolicdiscoveryoptimization}. Despite this reduced optimizer-state footprint, naive CXL achieves only 93\%--98\% of the baseline throughput, whereas CXL-aware allocation reaches 100\%--102\% (up to a 9\% improvement over naive CXL). Similarly, Figure~\ref{fig:ablation_studies}(c) explores a Mixture-of-Experts (MoE) model using \textbf{LFM2.5-8B-A1B}, which contains 8B total parameters but only about 1B active parameters~\cite{liquidai2026lfm258ba1b}. Although MoE execution alters the ratio of active parameters, the data still fundamentally separates into latency-sensitive optimizer states and latency-tolerant GPU transfer data. Consequently, naive CXL achieves 89\%--93\% of the baseline, whereas CXL-aware allocation reaches 98\%--99\%. Finally, Figure~\ref{fig:ablation_studies}(d) evaluates an encoder-decoder architecture using \textbf{T5Gemma 2 4B-4B}~\cite{zhang2025t5gemma2}, where CXL-aware allocation sustains 100\%--103\% of baseline throughput, up to a 12\% improvement over naive CXL.

Across variations in hardware platforms, optimizers, and model architectures, these results demonstrate that CXL-aware allocation generalizes effectively because it targets a fundamental, stable system-level property: CPU optimizer data are inherently latency-sensitive, while most offloaded GPU-transfer data are comparatively latency-tolerant.

\vspace{-0.15in}
\subsection{Comparison with OS-Level Transparent Tiering}\label{sec:eval:tiered_memory}

\begin{figure}[t]
    \centering
    \begin{minipage}{0.24\textwidth}
        \centering
        \includegraphics[height=1.23in]{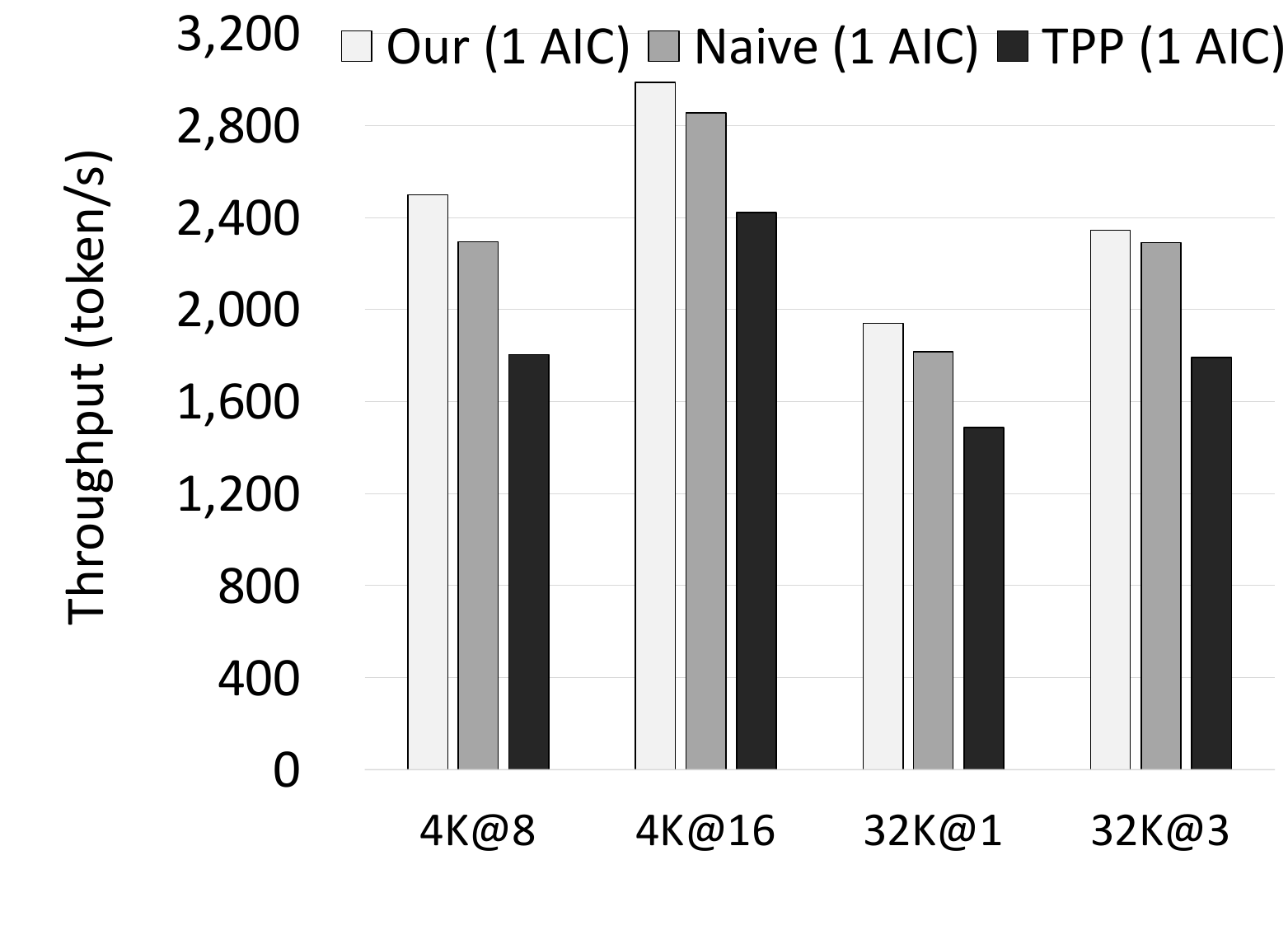}
        \caption{Throughput comparison against transparent tiered memory using TPP with AutoNUMA.}
        \label{fig:tpp_ablation}
    \end{minipage}\hfill
    \begin{minipage}{0.24\textwidth}
        \centering
        \includegraphics[height=1.23in]{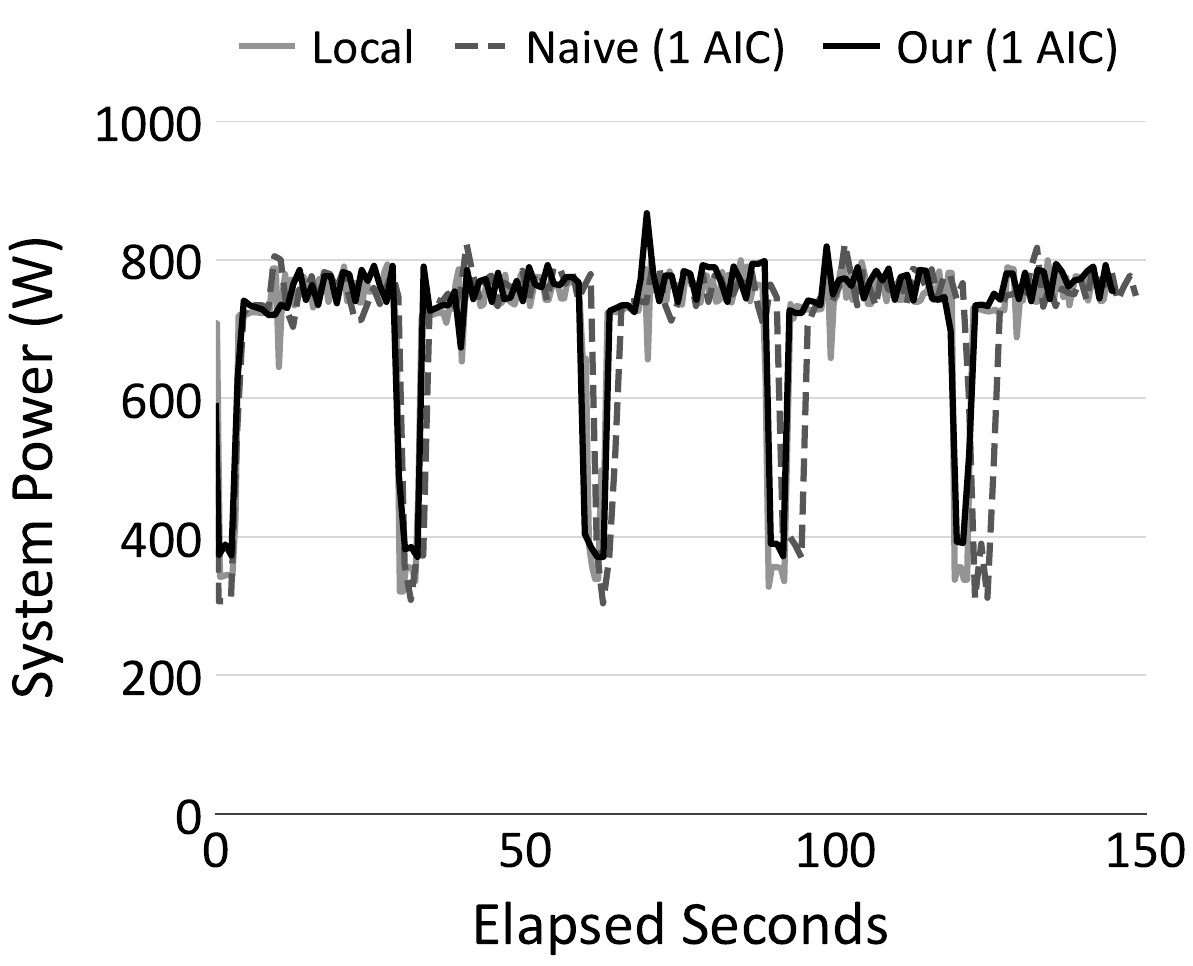}
        \caption{System power traces comparison for five fine-tuning iterations across different strategies.}
        \label{fig:energy_analysis_trace}
    \end{minipage}
    \vspace{-0.05in}
\end{figure}

This section compares CXL-aware allocation against transparent tiered memory mechanisms utilizing TPP~\cite{tpp2023} and AutoNUMA~\cite{autonuma_lwn}. Transparent tiering is traditionally attractive because it leverages CXL-attached memory capacity without requiring modifications to the training framework. Unlike the proposed application-aware approach, which uses tensor semantics to dictate placement at allocation time, TPP with AutoNUMA relies on OS-level page-access behavior to identify hot pages and dynamically migrate them between memory tiers. However, CPU-offloaded LLM fine-tuning exhibits two fundamental properties that render this OS-level mechanism ineffective. First, the workload lacks a clean hot/cold page separation. Offloaded parameters, gradients, optimizer states, and checkpointed activations are all touched during every training iteration. Consequently, most pages appear "hot" to an OS-level migration policy, even though their actual latency sensitivities differ substantially. Transparent tiering therefore cannot reliably anchor latency-critical optimizer data in local DRAM while leaving latency-tolerant transfer data in CXL-attached memory. Second, efficient CPU--GPU transfers strictly require CUDA pinned host memory. On the evaluated platform, pinned memory achieves approximately 50~GiB/s over PCIe x16, whereas unpinned memory throughput falls below 15~GiB/s. Because dynamic page migration is fundamentally incompatible with page-locked pinned buffers, TPP is forced to use unpinned memory, which directly bottlenecks the FWD and BWD phases.

The empirical results in Figure~\ref{fig:tpp_ablation} confirm these architectural limitations, with TPP throughput normalized to CXL-aware allocation across identically configured workloads (Section~\ref{sec:eval:ablation_on_5090}). Across all tested settings, TPP with AutoNUMA performs significantly worse than both naive CXL interleaving and CXL-aware allocation, achieving only 52.1\%--81.1\% of CXL-aware throughput at a 4K context length, and 76.1\%--76.7\% at 32K. The primary FWD/BWD penalty stems directly from the unpinned transfers: for a 4K context with a batch size of 8, TPP increases FWD+BWD latency to 14.46~s (compared to 10.81~s for naive CXL and 10.45~s for CXL-aware allocation). Furthermore, the STEP phase exposes the secondary limitation regarding page migration. Although STEP is not bottlenecked by the pinned-memory restriction, TPP still fails to distinguish latency-sensitive optimizer pages from latency-tolerant transfer pages because all offloaded data is accessed per iteration. As a result, STEP latency under TPP remains close to, or even worse than, naive CXL interleaving: 3.71~s versus 3.47~s at 4K context (batch size 8), and 4.07~s versus 3.52~s at 32K context (batch size 1). Ultimately, these results demonstrate that transparent tiering is insufficient for CPU-offloaded LLM fine-tuning; effective CXL-attached memory expansion strictly requires application-aware placement that proactively separates latency-sensitive CPU optimizer data from bandwidth-oriented GPU-transfer data.

\vspace{-0.15in}
\subsection{System Power Analysis}\label{sec:eval:energy_pattern_analysis}
To understand how the proposed allocation strategy affects total system power behavior, this section examines Uninterruptible Power Supply (UPS) power logs sampled at 0.5-second intervals. The comparison covers the local-DRAM baseline, naive CXL interleaving, and CXL-aware allocation across various context lengths and batch sizes. To ensure an accurate local-DRAM baseline, the CXL AIC is physically removed from the system to avoid capturing idle device power. Figure~\ref{fig:energy_analysis_trace} shows the total system power traces across five fine-tuning iterations for a representative long-context configuration, and Table~\ref{tab:energy_phase_power} reports the mean power of each phase for each configuration. Each iteration consists of the FWD, BWD, and STEP phases. Because the UPS readings are too coarse to resolve the short FWD phase, the two GPU-intensive phases are reported together as FWD+BWD. Across all three strategies, the traces exhibit the same recurring pattern: power consumption rises during the FWD and BWD passes and drops during the STEP phase. This behavior indicates that total system power is dominated by GPU and CPU computation rather than the memory-placement policy itself. As shown in Table~\ref{tab:energy_phase_power}, the system draws 719--776 W during FWD+BWD, whereas it drops to 387--477 W during the lower-power STEP phase. Within each configuration, the three strategies differ by at most 27 W during FWD+BWD, and their STEP differences show no consistent direction. Ultimately, these results demonstrate that CXL-aware allocation does not alter the instantaneous system power profile.

\begin{table}[h]
    \centering
    \scriptsize
    \setlength{\tabcolsep}{3pt}
    \vspace{-0.2in}
    \caption{Mean system power (W) of each phase across configurations.}
    \label{tab:energy_phase_power}
    \begin{tabular}{@{} l cc cc cc @{}}
        \toprule
        & \multicolumn{2}{c}{\textbf{Local}} & \multicolumn{2}{c}{\textbf{Naive (1 AIC)}} & \multicolumn{2}{c}{\textbf{Our (1 AIC)}} \\
        \cmidrule(lr){2-3} \cmidrule(lr){4-5} \cmidrule(l){6-7}
        \textbf{Config.} & FWD+BWD & STEP & FWD+BWD & STEP & FWD+BWD & STEP \\
        \midrule
        4K@8 & 719 & 409 & 728 & 387 & 746 & 453 \\
        4K@16 & 752 & 466 & 776 & 477 & 759 & 440 \\
        32K@1 & 721 & 426 & 727 & 451 & 732 & 437 \\
        32K@2 & 748 & 439 & 757 & 448 & 759 & 443 \\
        \bottomrule
    \end{tabular}
    \vspace{-0.15in}
\end{table}

\vspace{-0.15in}
\subsection{Performance Stability Across Multiple Runs}\label{sec:eval:stability}
Table~\ref{tab:tpp_variance} reports the latency stability for the Qwen2.5-7B workloads evaluated in the transparent-tiering comparison in Section~\ref{sec:eval:tiered_memory}. Each configuration was executed five times, and stability is quantified using the coefficient of variation (CV), defined as the standard deviation divided by the mean latency, and the normalized range, defined as $(\max-\min)/\text{mean}$.
Across all tested context-length and batch-size settings, the local-DRAM baseline, naive CXL interleaving, and CXL-aware allocation maintain a CV below 0.31\% and a normalized range below 0.91\%. In contrast, TPP with AutoNUMA exhibits substantially higher variability, with CVs ranging from 2.32\% to 7.87\% and normalized ranges from 6.76\% to 21.16\%. This instability arises because tiering depends on dynamic page migration and unpinned transfers, the timing of which can fluctuate across runs. By contrast, CXL-aware allocation fixes tensor placement at initialization and utilizes page-locked pinned memory for CPU--GPU transfers. These results confirm that this study improves average throughput while strictly maintaining execution-time predictability.

\begin{table}[h]
    \centering
    \scriptsize
    \setlength{\tabcolsep}{2.5pt}
    \vspace{-0.2in}
    \caption{Multiple run latency stability.}
    \label{tab:tpp_variance}
    \begin{tabularx}{\columnwidth}{@{} >{\RaggedRight}X c c @{}}
        \toprule
        \textbf{Configuration} & \textbf{CV Range (\%)} & \textbf{Norm. Range (\%)} \\
        \midrule
        local-DRAM baseline & 0.08--0.31 & 0.21--0.91 \\
        naive CXL & 0.05--0.29 & 0.14--0.79 \\
        CXL-aware allocation & 0.04--0.24 & 0.11--0.70 \\
        TPP with AutoNUMA & 2.32--7.87 & 6.76--21.16 \\
        \bottomrule
    \end{tabularx}
    \vspace{-0.1in}
\end{table}

\vspace{-0.15in}
\subsection{Impact of Latency-Sensitivity Ordering}
\label{sec:eval:latency_sensitivity_order}

\begin{figure}[!t]
    \centering
    \subfloat[Misplacing compute parameters in CXL-attached memory]{\includegraphics[height=1.23in]{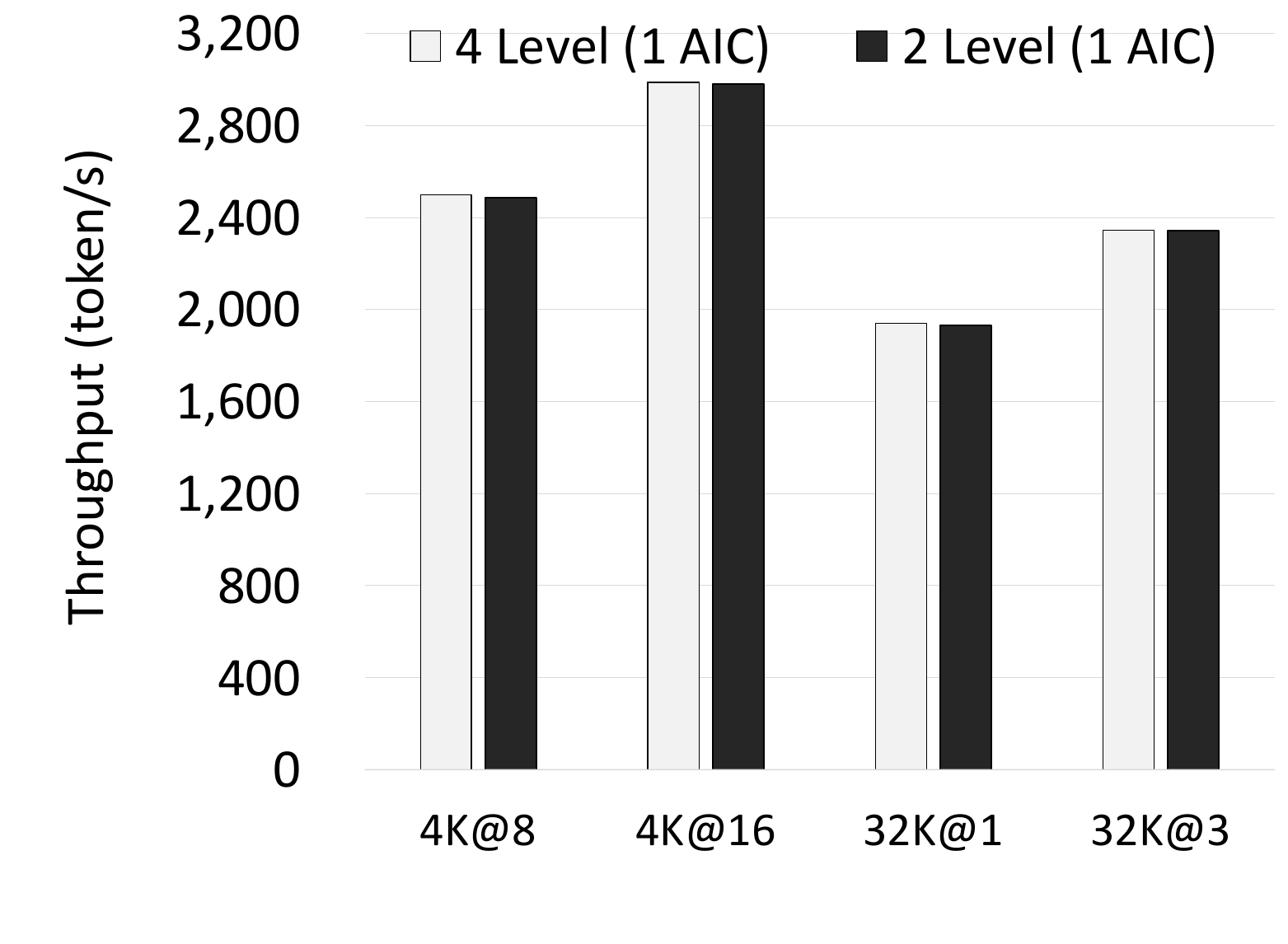}}
        \vspace{-0.01in}
    \subfloat[Misplacing checkpointed activations in CXL-attached memory]{\includegraphics[height=1.23in]{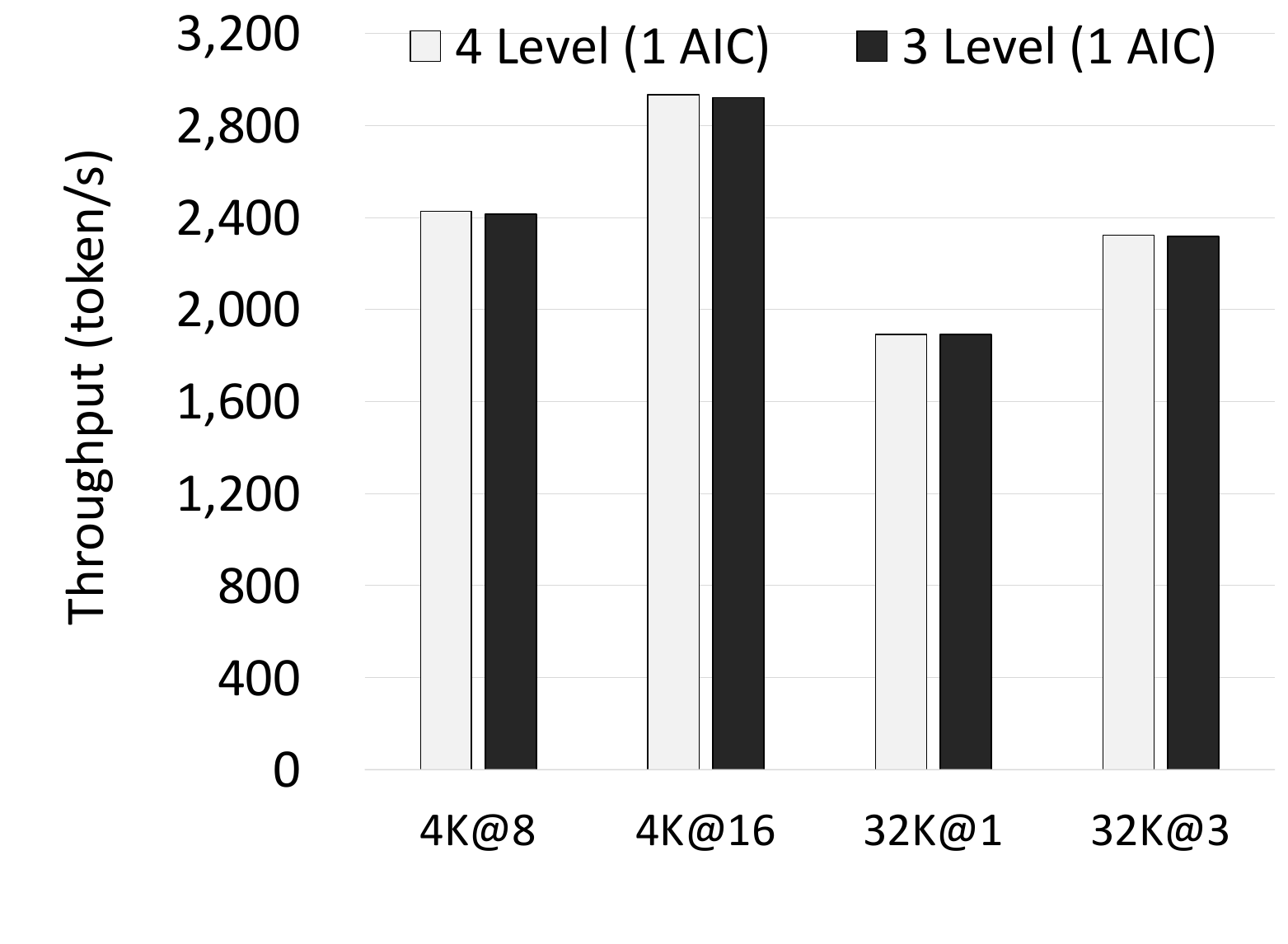}}
        \vspace{-0.01in}
    \caption{Ablation study of the latency-sensitivity ordering.}
    \vspace{-0.05in}
    \label{fig:ablation_studies_latency_order}
\end{figure}

The latency-sensitivity classification detailed in Section~\ref{sec:design:latency_sensitivity} is designed to enforce a deterministic allocation order between local DRAM and CXL-attached memory. This hierarchy prevents data with high sensitivity from being displaced to higher-latency CXL memory while less latency-sensitive data occupy local DRAM. This ordering is critical because CPU-offloaded fine-tuning generates multiple CPU-resident data groups with distinct access patterns. Specifically, CPU Adam data are accessed via tightly coupled, latency-sensitive CPU load--compute--store loops, whereas compute parameters, checkpointed activations, and compute gradients are primarily associated with bulk CPU--GPU transfers. To evaluate the practical impact of this ordering, an ablation study is conducted by intentionally violating the proposed hierarchy. Either compute parameters or checkpointed activations are deliberately misplaced into CXL-attached memory while keeping the rest of the placement unchanged.

Figure~\ref{fig:ablation_studies_latency_order}(a) illustrates the effect of misplacing compute parameters in CXL-attached memory. The STEP phase latency increases from an average of 2.65~s to 2.75~s (a 3.8\% slowdown), while the FWD and BWD phases remain nearly identical. This localized slowdown occurs because, after CPU Adam updates the \texttt{fp32} master parameters, the new values must be converted and copied into the \texttt{bf16} compute-parameter buffers. Offloading these buffers to CXL-attached memory increases the cost of this copy operation, directly prolonging the STEP phase. Figure~\ref{fig:ablation_studies_latency_order}(b) demonstrates the effect of misplacing checkpointed activations. The combined FWD+BWD latency increases from 19.21~s to 19.27~s for the 4K-context, batch-size-16 configuration, and from 39.14~s to 39.23~s for the 32K-context, batch-size-3 configuration. However, at smaller batch sizes, such as 4K with a batch size of 8 or 32K with a batch size of 1, the performance degradation is negligible. Because the checkpointed activation footprint is smaller under these conditions, the additional latency incurred by placing them in CXL-attached memory is effectively masked by asynchronous offloading and prefetching.

Overall, average throughput under these misplacement configurations retains 99.69\% of optimal performance, confirming that the primary benefit of CXL-aware allocation comes from isolating highly latency-sensitive CPU optimizer data from latency-tolerant GPU-transfer data. Nevertheless, GPU-transfer groups are not entirely interchangeable; inverting the placement of compute parameters and checkpointed activations introduces measurable overhead under specific memory-pressure conditions. Because the allocator natively supports per-tensor placement without runtime overhead, maintaining this fine-grained ordering is strictly preferable to using a single coarse class, maximizing robustness across various context lengths, batch sizes, and memory-pressure regimes. While a finer-grained, layer-by-layer hierarchy could theoretically exploit further access-pattern differences, its practical return is limited and would only serve to increase memory fragmentation and framework engineering complexity. Therefore, the current four-level design preserves the simplicity of a latency-first greedy policy, providing a highly practical balance between performance benefits and system complexity.

\vspace{-0.1in}
\section{Related Works} \label{sec:related_works}
Offloading strategies have emerged as an effective means of breaking the GPU memory ceiling, enabling the training of models that would otherwise exceed on-device capacity by staging tensors in CPU DRAM or NVMe SSDs~\cite{ren2021zerooffloaddemocratizingbillionscalemodel,rajbhandari2021zeroinfinitybreakinggpumemory,huang2023elixirtrainlargelanguage,chen2025practicaloffloadingfinetuningllm,zeng2025autoheteautomaticefficientheterogeneous}. Among these, the ZeRO series~\cite{ren2021zerooffloaddemocratizingbillionscalemodel,rajbhandari2021zeroinfinitybreakinggpumemory} has become the most widely adopted, owing to continuous maintenance that fixes bugs, adds support for new models, and preserves interoperability with complementary optimizations such as Flash-Attention~\cite{dao2023flashattention2fasterattentionbetter} and Liger-Kernel~\cite{hsu2025ligerkernelefficienttriton}. Its reference implementation now underpins several popular training frameworks, such as Accelerate~\cite{accelerate}. Complementary work further broadens the design space of memory-centric training. Huang et al.~\cite{huang2023elixirtrainlargelanguage} tailor an offloading mechanism specifically for large-scale language models, while Chen et al.~\cite{chen2025practicaloffloadingfinetuningllm} evaluate pragmatic orchestration policies that coordinate CPU and GPU memory during fine-tuning. Zeng et al.~\cite{zeng2025autoheteautomaticefficientheterogeneous} extend the idea to fully heterogeneous environments, automatically balancing both compute and memory resources.

For CXL-attached memory, several tiered-memory systems (TMS) address the capacity challenge at the operating-system or hardware level. Representative examples include TPP~\cite{tpp2023}, which classifies pages as hot or cold for placement, and NOMAD~\cite{nomad}, which employs transactional page migration for asynchronous data movement. Other advanced systems, such as Colloid~\cite{colloid}, dynamically balance traffic to equalize effective latency, while M5~\cite{m52025masteringpagemigration} embeds hardware trackers in the CXL controller to provide fine-grained access statistics. Although these general-purpose TMS designs operate transparently without requiring application modifications, their workload-agnostic nature can lead to suboptimal performance for specialized applications such as LLM fine-tuning. Recent CXL-based LLM research primarily targets inference, such as KV-cache storage~\cite{cxlmemoryexpansionforkvcachestorage} and parameter offloading~\cite{kim2025lia}. To our knowledge, this work is one of the pioneer application-aware solutions to address CXL latency specifically for CPU-offloaded LLM fine-tuning.

\vspace{-0.1in}
\section{Conclusion}\label{sec:conclusion}

While the fundamental DRAM–CXL performance gap has been clearly revealed by previous studies, this work demonstrates that CPU-offloaded long-context fine-tuning requires tensor-level placement. Because current PyTorch and OS NUMA interfaces cannot express fine-grained placement priorities, they fail to distinguish between highly latency-sensitive CPU optimizer data and GPU-transfer tensors that can effectively mask CXL latency. To address the latency challenge of CXL-attached memory relative to local DRAM during long-context LLM fine-tuning, particularly the performance degradation observed in CPU-intensive computations, this study identifies a fundamental limitation in existing deep learning frameworks: the lack of fine-grained control over memory allocation across heterogeneous memory systems. To overcome this limitation, two complementary approaches are proposed. First, a fine-grained memory allocation extension to PyTorch is developed to provide tensor-level control over system memory allocation policies, enabling fine-grained tensor management and facilitating future research with CXL-attached memory. Second, a CXL-aware memory allocator is introduced, which strategically places latency-sensitive data in local DRAM and latency-tolerant data in CXL-attached memory based on their access patterns. Together, these optimizations substantially mitigate the performance drawbacks of CXL-attached memory, consistently outperforming naive CXL with up to 21\% improvement across evaluated scenarios. In the dual-AIC configuration, CXL-aware allocation achieves near-baseline throughput, narrowing the gap with the local-DRAM baseline to within 1\%, thereby demonstrating that CXL-attached memory can serve as a practical and high-performance solution for long-context LLM fine-tuning.

\vspace{-0.1in}
\bibliographystyle{IEEEtran}
\bibliography{references}

\end{document}